\documentclass[sigconf]{acmart}

\usepackage[yyyymmdd]{datetime}

\usepackage{tabularx}
\usepackage{dcolumn} 
\newcolumntype{d}[1]{D{.}{.}{#1}}

\usepackage{subcaption}
\usepackage{todonotes}
\let\xtodo\todo
\renewcommand{\todo}[1]{\xtodo[inline,color=green!50]{#1}}

\AtBeginDocument{%
  }

\copyrightyear{2026}
\acmYear{2026}
\setcopyright{cc}
\setcctype{by}
\acmConference[CHI '26]{Proceedings of the 2026 CHI Conference on Human Factors in Computing Systems}{April 13--17, 2026}{Barcelona, Spain}
\acmBooktitle{Proceedings of the 2026 CHI Conference on Human Factors in Computing Systems (CHI '26), April 13--17, 2026, Barcelona, Spain}
\acmDOI{10.1145/3772318.3790523}
\acmISBN{979-8-4007-2278-3/2026/04}

\begin{document}

\title[Anticipation  Before Action]{Anticipation Before Action: EEG-Based Implicit Intent Detection for Adaptive Gaze Interaction in Mixed Reality}

\settopmatter{authorsperrow=4}

\author{Francesco Chiossi}
\orcid{0000-0003-2987-7634}
\affiliation{
  \institution{LMU Munich}
  \city{Munich}
  \country{Germany}
}
\email{francesco.chiossi@ifi.lmu.de}

\author{Elnur Imamaliyev}
\orcid{0009-0005-2331-0073}
  \affiliation{%
    \institution{Carl von Ossietzky Universität Oldenburg}
    \city{Oldenburg}
    \country{Germany}}
\email{elnur.imamaliyev@uni-oldenburg.de}

\author{Martin G. Bleichner}
\orcid{0000-0001-6933-9238}
\affiliation{%
  \institution{Carl von Ossietzky Universität Oldenburg}
  \city{Oldenburg}
  \country{Germany}
  }
\email{martin.georg.bleichner@uol.de}

\author{Sven Mayer}
\orcid{0000-0001-5462-8782}
\affiliation{%
  \institution{TU Dortmund University}
  \city{Dortmund}
  \country{Germany}}
\affiliation{%
  \institution{Research Center Trustworthy Data Science and Security}
  \city{Dortmund}
  \country{Germany}}
\email{info@sven-mayer.com}

\renewcommand{\shortauthors}{Chiossi et al.}

\begin{abstract}
Mixed Reality (MR) interfaces increasingly rely on gaze for interaction, yet distinguishing visual attention from intentional action remains difficult, leading to the Midas Touch problem. Existing solutions require explicit confirmations, while brain–computer interfaces may provide an implicit marker of intention using Stimulus-Preceding Negativity (SPN).
We investigated how Intention (Select vs. Observe) and Feedback (With vs. Without) modulate SPN during gaze-based MR interactions. 
During realistic selection tasks, we acquired EEG and eye-tracking data from 28 participants.
SPN was robustly elicited and sensitive to both factors: observation without feedback produced the strongest amplitudes, while intention to select and expectation of feedback reduced activity, suggesting SPN reflects anticipatory uncertainty rather than motor preparation.
Complementary decoding with deep learning models achieved reliable person-dependent classification of user intention, with accuracies ranging from 75\% to 97\% across participants.
These findings identify SPN as an implicit marker for building intention-aware MR interfaces that mitigate the Midas Touch.
\end{abstract}





\maketitle

\section{Introduction}

Mixed reality (MR) interfaces are advancing usability and accessibility, with devices like Apple's Vision Pro offering gaze and hand-based interactions to support more natural control in spatial computing environments~\cite{pfeuffer2017gaze, lystbaek2024hands, jinwook2025pinch}. Despite these advances, reliably distinguishing between visual attention and intentional interaction remains a persistent usability challenge, commonly referred to as the Midas Touch problem~\cite{jacob1990what}. This issue arises when systems misinterpret mere looking as deliberate action, often leading to unintended commands~\cite{burnham2025effects}. Because gaze is inherently continuous and exploratory, treating it as direct input can compromise the fluidity and reliability of gaze-based control. Dwell-based techniques, where gaze fixations are interpreted as selections after a fixed duration~\cite{penkar2012designing}, are especially prone to false positives~\cite{narkar2024gazeintent,yan2020head}, leading to unintentional selections that disrupt user experience~\cite{tian2025amplitude}. Thus, there is a lack of implicit, intention-sensitive mechanisms that can reliably disambiguate whether a user intends to interact with a gaze target or is simply observing.

Previous work has attempted to mitigate gaze ambiguity by introducing external confirmation signals, such as hand gestures~\cite{kyto2018pipointing, mutasim2021pinch}, speech commands~\cite{reiter2022look, whithlock2018interacting}, or blink-based triggers~\cite{lu2021itext}. More recently, brain-computer interface (BCI) approaches have shown promise in inferring cognitive states in MR~\cite{nwagu2023eeg}. Notably, \citet{reddy2024towards} demonstrated that the Stimulus-Preceding Negativity (SPN), a slow cortical potential that builds up before anticipated events~\cite{vanBoxtel2004cortical}, can serve as an electrophysiological (EEG) correlate of intent in controlled Virtual Reality (VR) environments. While this work validated SPN for target selection tasks, questions remain about how SPN responds to key interactive variables, such as user intention and system feedback, across more ecologically valid gaze-based tasks. Understanding how intention and feedback jointly shape anticipatory neural responses would support the design of multimodal interfaces that can distinguish purposeful gaze from casual observation. This knowledge could enable hands-free, confirmation-free interactions that adapt to users' internal states in real time.

In this work, we systematically investigate how user intention (\textit{Select} vs. \textit{Observe}) and feedback (\textit{Feedback} vs. \textit{No Feedback}) jointly influence anticipatory EEG activity during gaze-based MR interactions. This dual manipulation is important because previous studies conflated intention with system response, making it unclear whether SPN reflects internal goals or external feedback expectations. Unlike previous studies that focused on narrow selection paradigms~\cite{shishkin2016eeg, kotani2025stimulus}, we embed our study within diverse realistic scenarios, including app selection, UI controls, and media playback in MR. Using EEG, we test whether SPN can differentiate intention types and whether it reflects internal anticipation rather than merely feedback presence. Moreover, we assess whether such EEG signatures can be classified using machine learning models for practical MR applications.

Our results demonstrate that SPN amplitude was shaped by both intention and feedback, with observation without feedback eliciting the strongest anticipatory activity. Overall, the intention to act and the expectation of feedback reduced SPN amplitude, suggesting that in MR contexts SPN reflects anticipatory uncertainty rather than motor preparation. This establishes SPN as a consistent anticipatory marker during realistic MR tasks and refines its interpretation beyond motor intent.

We also explored whether intention can be decoded directly from EEG signals using deep learning models in a person-dependent setup. Classification accuracies reached up to 97\%, with \textit{EEGInceptionERP} performing best, showing that anticipatory brain activity contains discriminative information about user goals. This demonstrates the feasibility of intention decoding in MR, while also pointing toward the need for personalized calibration.

Taken together, these findings contribute in four ways: (1) they confirm that SPN is robustly elicited across ecologically valid MR interactions, (2) they show that SPN indexes anticipatory uncertainty rather than motor preparation, shaped by the joint influence of intention and feedback, (3) they demonstrate that intention can be decoded from anticipatory EEG using deep learning in a person-dependent setup, and (4) they outline implications for adaptive MR design, where SPN patterns could help reduce false activations (Midas Touch), guide feedback timing, and be combined with other modalities (e.g., pupil size, electrodermal activity (EDA)) to build intention-aware and uncertainty-sensitive interfaces.

\section{Related Work}
In the following, we first review gaze-based interaction techniques in MR and their associated usability challenges. We then examine prior work on disambiguating user intent in gaze-driven interfaces, including both explicit and implicit approaches. Finally, we discuss EEG correlates of intention, focusing on the SPN component as a candidate signal for adaptive interaction.

\subsection{Gaze-Based Interaction in Mixed Reality}

Gaze is valued for its intuitive, natural, and hands-free characteristics. It enables users to interact without holding physical controllers, reducing physical load and aligning interaction closely with attention. This makes it particularly well-suited for MR scenarios where mobility, context awareness, and low-effort input are common requirements. 

Thus, it emerged as a prominent input modality in MR. A wide range of gaze-based techniques have been proposed, with dwell-based selection being the most commonly used.
In this method, a user selects a target by fixating on it for a predefined duration, an approach introduced by \citet{jacob1990what}. Other systems, such as that of \citet{starker1990gaze}, used dwell not for selection but as a signal of user interest, triggering additional information after a fixed delay.

More recent gaze-based methods include raycasting \cite{chen2023gaze, gabel2024guiding,gabel2023redirecting}, where gaze or head orientation \cite{hendrikson2020head} guides a virtual pointer, and cursor, based overlays, which provide visual feedback on gaze position and potential targets.

Despite its advantages, gaze-based interaction still faces persistent usability issues. Chief among them is the Midas Touch problem, where systems struggle to distinguish between passive viewing and intentional action, often triggering unintended commands \cite{jacob1990what}. Another significant challenge is limited gaze precision, particularly in the periphery, where spatial targeting is compromised by the eye’s lower visual acuity and inaccuracies in head-mounted eye tracking \cite{koulieris2019neareye,plopski2022eye}. 
This makes fine-grained selection difficult in cluttered MR environments. A third issue is calibration drift, where eye-tracking accuracy degrades over time due to slippage, lighting changes, or user movement, often requiring frequent recalibration that interrupts the user experience~\cite{niehorster2020slippage, plopski2022eye}. In addition, factors such as occlusions, eyewear, or individual physiology can further reduce tracking robustness~\cite{plopski2022eye}.
Together, these challenges highlight the need for more robust and selection-sensitive mechanisms that can enhance the reliability and usability of gaze-based interaction in MR.


\subsection{Disambiguating Intent in Gaze-Driven Interfaces}

A key step toward more robust gaze-based interfaces is the ability to disambiguate between when a user is simply looking at content and when they intend to act on it. To address this, many systems have introduced explicit confirmation techniques that combine gaze with auxiliary modalities. Examples include voluntary or half-eye blinks \cite{lu2021itext, lee2014ar}, speech commands \cite{whithlock2018interacting}, head  \cite{pfeuffer2017gaze, pfeuffer2024gazepinch} and body movements \cite{klamka2015lookpedal}. 
Here, hand gestures, such as the pinch gesture \cite{pfeuffer2017gaze}, have proven particularly effective, and additional gaze-hand alignment techniques, such as Gaze \& Handray \cite{gabel2024guiding} or Gaze \& Finger \cite{wagner2023fitts}, confirm target selection by aligning hand rays or fingers with gaze \cite{horstmann2005target, mrotek2007target}.

While these multimodal approaches mitigate the “Midas touch” problem, they also introduce drawbacks. Gestural interfaces often cause physical fatigue, making them difficult to sustain during long sessions \cite{mutasim2021pinch, park2024impact}. Gaze interactions themselves may lead to eye fatigue or strain \cite{hirzle2020survey}, although studies suggest less strain when gaze is limited to pointing rather than being used for both pointing and selection \cite{wagner2023fitts, pfeuffer2020empirical, zavichi2025gazehand}. Similarly, blink- or vergence-based techniques can be uncomfortable and demand additional cognitive effort \cite{kirst2016verge, chiossi2024mind}. 

To overcome these limitations, researchers have begun to investigate implicit, intent-sensitive signals. \citet{sharma2024distinguishing} demonstrated that combining EEG with eye tracking enables classification of target versus non-target fixations during free visual search in realistic, cluttered environments. Their multimodal model achieved over 80\% cross-user accuracy, substantially outperforming earlier ERP-based methods. Their findings show that intent can be disambiguated not only through external confirmation gestures, but also through implicit neural and ocular markers of goal-directed fixations. This neuro-adaptive approach highlights a promising path forward, even though its reliance on EEG still limits its everyday practicality in MR. Along similar lines, multimodal fusion of eye tracking with EEG has been shown to discriminate navigational from informational search intentions \citep{park2014}, while EEG phase synchrony features further improved recognition of these implicit states \citep{kang2015implicit}. Extending to VR, time-domain EEG and electromyographic (EMG) signals can reveal idle versus pre-movement states, enabling zero-lag interaction \citep{gehrke2025}.  Together, these works point toward intent-sensitive interfaces that anticipate rather than react to user actions.


\subsection{EEG Correlates of Intention}

Intention is broadly defined as a conscious representation of an action that includes a goal to be achieved, serving as a mental prefiguration of future behavior \cite{delnatte2023freewill}. It is often described as a perception without causal power, meaning it reflects the planning or anticipation of action rather than its direct execution \cite{reddy2024towards}. Here, intention is tied to the sense of agency: when the outcome of an action matches one's internal intention, it reinforces the subjective experience of authorship and control over that action \cite{villa2025agency}. In human–computer interaction (HCI), intention is necessary to enable systems to interpret user goals \cite{jackson1997behavioral}. It is typically operationalized as a user's internal commitment to perform a specific action, such as selecting a UI element, issuing a command, or interacting with content. 

Here, electrophysiological signals offer a promising pathway to implicit intent detection \cite{zhang2020making, reddy2024towards}. Within EEG research, three components have been most consistently associated with intention: the Readiness Potential (RP), the Lateralized Readiness Potential (LRP), and the Stimulus-Preceding Negativity (SPN).

The RP is a slow negative fluctuation emerging up to two seconds before voluntary movement \cite{deecke1969readiness, libet1983preparation}. It originates in the supplementary motor areas and becomes strongest contralateral to the acting hand shortly before movement onset \cite{shibasaki2006bereitschaftspotential}. Because RP reliably precedes both the execution of movement and the subjective awareness of intention, it has been central to debates on free will \cite{schurger2021readiness}. Despite these controversies, RP is widely accepted as a robust signal of subconscious motor preparation \cite{shibasaki2006bereitschaftspotential}, though recent evidence suggests it may reflect more general decision-related or anticipatory processes rather than pure motor planning \cite{alexander2016readiness}. The LRP refines this further by indexing effector-specific preparation over motor cortices \cite{syrov2024visuomotor}, reflecting whether the left or right hand will act. Both RP and LRP, however, are bound to motor execution and thus less suited to interaction paradigms where intentions may not culminate in movement.

The SPN captures a fundamentally different aspect of anticipation \cite{brunia2012negativeslow}. It is a slow negative potential that builds up in the interval before an expected, task-relevant stimulus \cite{damen1994spn}. Critically, unlike RP and LRP, SPN is a \textit{non-motor} anticipatory component: it emerges in paradigms without movement, with participants simply awaiting informative stimuli \cite{vanboxtel1994motor}. Converging evidence establishes that SPN specifically indexes \textit{anticipatory uncertainty} rather than general expectancy. \citet{catena2012brain} demonstrated that SPN amplitude is larger for probabilistically unpredictable outcomes than for predictable ones, with source localization implicating prefrontal areas related to uncertainty processing. \citet{tanovic2019anticipating} showed that an uncertain threat (50\% shock probability) elicits significantly stronger SPN than a certain threat (100\%) or safety (0\%), establishing SPN as specifically sensitive to uncertainty beyond valence or arousal. \citet{walentowska2018relevance} further demonstrated that relevance and uncertainty jointly modulate SPN, with maximal amplitudes when outcomes are both uncertain and personally significant.

These empirical findings align with contemporary neurocognitive theories based on predictive coding and the free energy principle \cite{clark2013whatever}. Under these frameworks, the brain continuously generates predictions about sensory input and minimizes prediction errors by updating internal generative models \cite{friston2009predictive}. \citet{friston2009predictive} proposed that attention involves inferring the \textit{precision} (inverse uncertainty) of predictions, i.e.,allocating processing resources where predictions are most uncertain to prepare for potential prediction errors. SPN can be understood as an electrophysiological signature of this anticipatory uncertainty monitoring: when upcoming events cannot be reliably predicted, the brain allocates attentional resources to track prediction error and uncertainty \cite{catena2012brain, feldman2010attention}, manifesting as increased SPN amplitude. Conversely, when outcomes are predictable, either because contingencies are well-learned or because intentions clarify likely consequences, anticipatory uncertainty reduces, and SPN amplitude attenuates. Classically, SPN has been observed in tasks where participants anticipate motivationally salient events, such as gambling outcomes \cite{kotani2003spn}, performance feedback \cite{protzak2013passivebci}, or the arrival of an informative cue \cite{brunia2012negativeslow}. In these settings, SPN reflects the allocation of cognitive resources toward upcoming information, scaling with the degree of uncertainty about what will follow \cite{megias2018electroencephalographic}.

This makes SPN particularly relevant for MR interfaces, where the challenge is distinguishing observation from intention under uncertainty. When users gaze at interface elements, their internal mental model determines whether they can predict what will follow: intentional selection establishes clear action-outcome predictions, reducing anticipatory uncertainty; passive observation, especially without feedback, creates maximal uncertainty about system response, heightening anticipatory monitoring demands. While prior work in VR shows SPN can separate intent-to-select from passive viewing \cite{reddy2024towards}, it remains unclear how SPN behaves in realistic MR tasks where both intention and feedback jointly shape anticipatory uncertainty. Our study addresses this gap by testing SPN in an Intent × Feedback design within ecologically valid MR scenarios, advancing its use as a marker for adaptive, intent-sensitive interaction.

\section{Hypotheses}

Building on prior work establishing SPN as a marker of anticipatory processing \cite{reddy2024towards, brunia2012negativeslow} and evidence linking SPN to uncertainty monitoring \cite{catena2012brain, tanovic2019anticipating, walentowska2018relevance}, we test the following hypotheses:

\begin{itemize}
  \item \textbf{H1:} SPN will be reliably elicited during gaze-based MR interactions across all experimental conditions, extending prior findings from simplified laboratory tasks to ecologically valid interface contexts.
  
  \item \textbf{H2:} SPN amplitude will be greater (more negative) in Observe than Select conditions. This prediction follows from the uncertainty account of SPN \cite{catena2012brain, feldman2010attention}: passive observation without established action-outcome contingencies creates greater uncertainty about what will follow, triggering enhanced anticipatory monitoring. In contrast, intentional selection establishes clear predictions about system response, reducing anticipatory uncertainty and thus SPN amplitude.
  
  \item \textbf{H3:} Intention and feedback will interact to modulate SPN amplitude. In Select trials, clear action intentions reduce baseline uncertainty, so feedback availability should have minimal additional impact on SPN. In Observe trials, baseline uncertainty is higher, so feedback presence should reduce SPN by resolving uncertainty about the system's response. This interaction would demonstrate that SPN tracks the degree of anticipatory uncertainty shaped by both internal goals and external information \cite{friston2009predictive, feldman2010attention}.
  
\end{itemize}

While these hypotheses address theoretical accounts of anticipation in MR, we also aim to examine the practical feasibility of decoding user intention directly from neural signals. Recent advances in deep learning applied to EEG suggest that classification-based approaches can move beyond theory-driven markers and support adaptive systems in practice \cite{lotte2018review, schirrmeister2017deepeeg, long2024detecting}. Thus, we formulate : 

\textbf{RQ:} To what extent can intention (\textsc{Select} vs. \textsc{Observe}) be decoded from EEG signals using deep learning models?

\section{User Study}

Previous research has linked the SPN to anticipatory processing in feedback and outcome tasks \cite{brunia2012negativeslow, kotani2003spn,protzak2013passivebci}  and, more recently, to intent detection in simplified VR settings \cite{reddy2024towards}. Yet, it remains unclear how SPN behaves in realistic MR tasks, where users alternate between observation and selection, i.e. \textsc{Intent} and where \textsc{Feedback} expectations shape anticipation. 


\subsection{Study Design}

We employed a within-participants experimental design. The independent variables were \textsc{Feedback} (two levels: Feedback / No Feedback) and \textsc{Intent} (two levels: Observe / Select). This yielded a $2 \times 2$ factorial structure. We add the Feedback conditions in the intent-to-observe scenario to address a key confounding factor: whether the SPN is tied to the user's anticipation of feedback or their intention to select \cite{reddy2024towards}.
To avoid learning effects, the order of conditions was counterbalanced across participants using a balanced Latin Williams square design with four levels~\cite{wang2009construction}.

For ecological validity, tasks were performed across three everyday MR scenarios where digital interface elements were overlaid onto the real-world environment viewed through the MR headset: (1) observing or selecting an \textit{app icon} in a digitally rendered launcher interface, (2) observing or selecting an \textit{icon} embedded within a virtual textual document, and (3) observing or selecting a \textit{control icon} to pause/play/stop a digitally presented video. All UI elements, feedback signals, and media content were computer-generated and spatially anchored in the participant's real-world visual field, while the background environment remained their actual physical surroundings. In the feedback conditions, participants received visual confirmation following the observation or selection action (e.g., highlighting the icon or app), whereas in the no-feedback conditions, no such confirmation was presented.

\begin{figure*}[htbp]
    \centering
    \includegraphics[width=\textwidth]{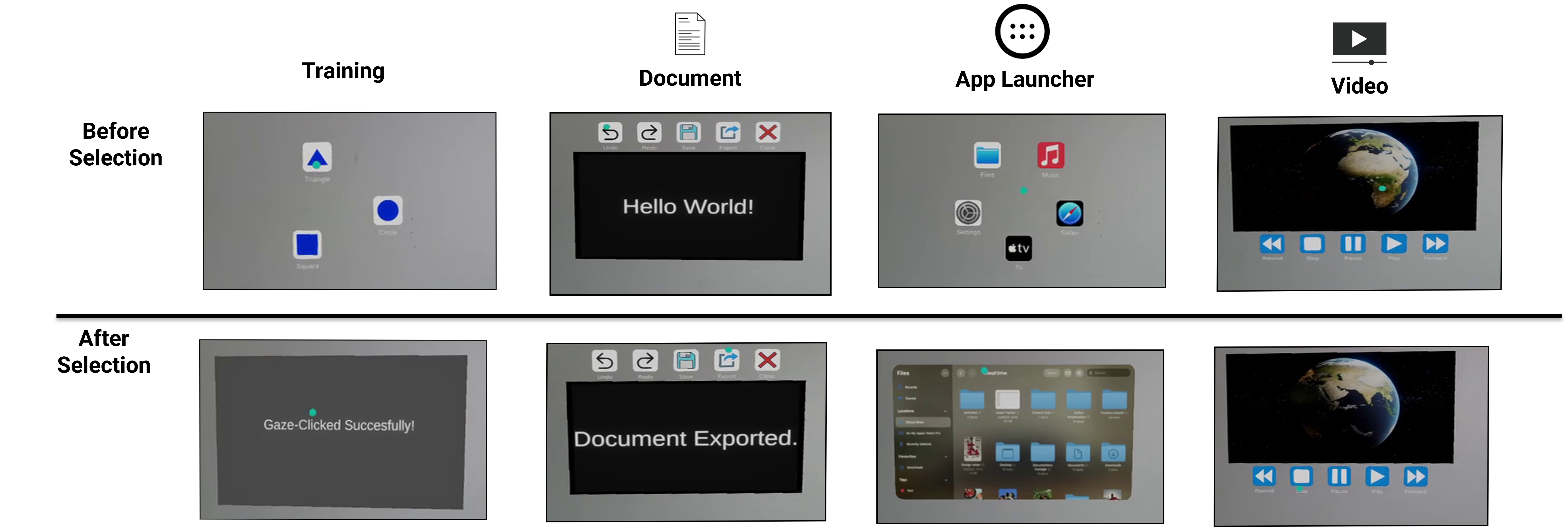}
    \caption{
    \textbf{Experimental scenarios illustrated with MR interfaces screenshots.} 
    Columns represent the four tasks (\textit{Training}, \textit{Document}, \textit{App Launcher}, \textit{Video}), 
    and rows show the interface states \textit{Before Selection} (top) and \textit{After Selection} (bottom). 
    During training, participants interacted with neutral geometric shapes, while in the main scenarios they 
    performed gaze-based interactions in everyday MR contexts such as editing text, launching applications, 
    browsing files, and controlling video playback. Feedback was presented immediately after gaze-based selections.
    }
    \label{fig:scenarios}
\end{figure*}

\subsection{Independent Variables}
We employed a within-participants experimental design with two independent variables: \textbf{Intention} and \textbf{Feedback}. Together, these factors produced four conditions: \textit{Select–With Feedback}, \textit{Select–Without Feedback}, \textit{Observe–With Feedback}, and \textit{Observe–Without Feedback}.

\subsubsection{Intention}
The first independent variable manipulated whether participants actively interacted with the interface or passively monitored it. 

In the \textit{Select} condition, participants were instructed to fixate on a target icon \textit{to select it}, triggering a specific UI action (e.g., launching an app, modifying a document, controlling video playback). In the \textit{Observe} condition, participants fixated on the icon \textit{to observe it}, with explicit instruction that no action would be triggered. Both conditions required 750 ms fixation.

Although oculomotor behavior was identical, these instructions induced different \textit{internal goal states}. This manipulation builds on foundational work by \citet{posner1980orienting}, who established that symbolic cues can shift covert attention and produce measurable cognitive effects despite participants maintaining central fixation, thus demonstrating that internal attentional states can be dissociated from overt behavior.
Subsequent research has consistently shown that task instructions create distinct anticipatory brain states despite matched behavioral outputs: endogenous attention cuing produces different neural signatures when participants are cued to ``attend left'' versus ``attend right'' despite maintaining central fixation \citep{kastner2000mechanisms}; task-set instructions (``respond to color'' vs. ``respond to shape'') elicit preparatory differences before stimuli appear \citep{sakai2008task}; and observing actions with intent to imitate versus passively watching yields distinct neural patterns despite identical visual input \citep{iacoboni1999cortical}. Our manipulation extends this principle to gaze-based MR interaction, where gaze can function as an instrumental command (Select) or passive monitoring (Observe).

We reinforced this distinction through: (1) explicit trial-initial instructions stating the goal (``Fixate to select [icon]'' vs. ``Fixate to observe [icon]''), (2) ecological scenario embedding (app launcher, document editor, media player) where this distinction is naturalistic, and (3) training with neutral shapes where participants directly experienced differential consequences (action vs. no action), establishing contingency learning.

The 750 ms dwell threshold was based on \citet{reddy2024towards}, who demonstrated successful SPN-based intention detection at this timing in VR. This duration falls within the typical 500 - 1000 ms range for SPN elicitation \citep{shishkin2016eeg}.


\subsubsection{Feedback}
The second independent variable manipulated whether visual confirmation of fixation was provided. In the \textit{With Feedback} condition, a 500 ms pop-up feedback signal was displayed after fixation (in Observe trials) or immediately before the UI action (in Select trials). The 500ms duration was chosen to ensure reliable perception and neural activation \cite{kohrs2016delays} while maintaining interaction responsiveness at the threshold where longer delays impair user behavior \cite{liu2014effects}. This timing optimizes feedback salience for our MR environment without compromising ecological validity. In the \textit{Without Feedback} condition, no feedback was shown, and the trial progressed directly to the next phase.

\subsection{Task}
\label{sec:task}
Participants performed a gaze-based interaction task in MR, where virtual UI elements were superimposed on their physical environment through the headset.  At the beginning of each trial, an instruction screen indicated whether they should \textit{Select} or \textit{Observe} the upcoming icon. To increase ecological validity, trials were embedded in three everyday usage scenarios: (1) an application launcher, (2) a text document, and (3) a video player, see \autoref{fig:scenarios}. The real-world surroundings served as the visual backdrop, while all interactive components, including application icons, document interfaces, and video controls, were digitally rendered and spatially registered within the participant's field of view.

The experimental structure fully crossed \textsc{Intent} (Observe / Select) and \textsc{Feedback} (Feedback / No Feedback) with the three scenarios. Each participant therefore completed all four conditions in each scenario, resulting in twelve blocks in total. Before the main experiment, participants performed a short training phase to practice each condition. The order of experimental blocks was counterbalanced across participants to minimize learning and carry-over effects.

\subsubsection{Scenarios}
Each trial was embedded in a scenario that provided contextually meaningful icons. In the \textit{App scene}, participants viewed a grid of application icons resembling a mobile home screen, including \textit{Files}, \textit{Music}, \textit{Settings}, \textit{Safari}, and \textit{TV}. In the \textit{Document scene}, they saw a text-editing interface with the title “Hello World!” displayed in the document area, accompanied by toolbar icons such as \textit{Undo}, \textit{Redo}, \textit{Save}, \textit{Export}, and \textit{Close}. In the \textit{Video scene}, they viewed a video player showing a clip of planet Earth, with playback controls including \textit{Rewind}, \textit{Stop}, \textit{Pause}, \textit{Play}, and \textit{Fast Forward}. We deliberately chose a neutral Earth video to avoid eliciting affective responses, as the SPN has been shown to be sensitive to affective state \citep{poli2007spn}.

\subsubsection{Select Condition}
In the Select condition, participants were instructed to actively interact with the interface by triggering a gaze-click. After the fixation cross, an icon from the current scenario appeared on the screen. Participants fixated on the icon for 750 ms, after which the system registered the gaze as a selection. Depending on the feedback condition, either a 500 ms pop-up feedback was shown before the system after the action (with Feedback), or the action was executed immediately (no Feedback). 

The specific outcome of a selection depended on the active scenario. For example, in the \textit{App scene}, selecting the \textit{Safari} icon opened the Safari browser window. In the \textit{Document scene}, selecting the \textit{Undo} icon removed the displayed text “Hello World!” from the document. In the \textit{Video scene}, selecting the \textit{Play} icon initiated playback of the Earth video. These contextualized responses ensured that each selection had a clear and meaningful consequence.

\subsubsection{Observe Condition}
In the Observe condition, participants were instructed to passively monitor the display without initiating an action. After the fixation cross, an icon from the current scenario appeared, and participants fixated on it for the same 750 ms dwell time. In the With Feedback trials, a 500-ms pop-up feedback was shown after fixation, whereas in the Without Feedback trials, no feedback was provided. No application, document, or video was launched in these trials, as the task only required sustained fixation on the presented icon.

\subsection{Procedure}
\label{sec:procedure}

Upon arrival, participants were informed about the study procedure and provided written informed consent. The experimenter then set up the water-based EEG recording. Participants wore the Varjo XR-4 headset and completed a five-point eye-tracking calibration before starting the experiment.

\begin{figure*}[t]
    \centering
    \includegraphics[width=\linewidth]{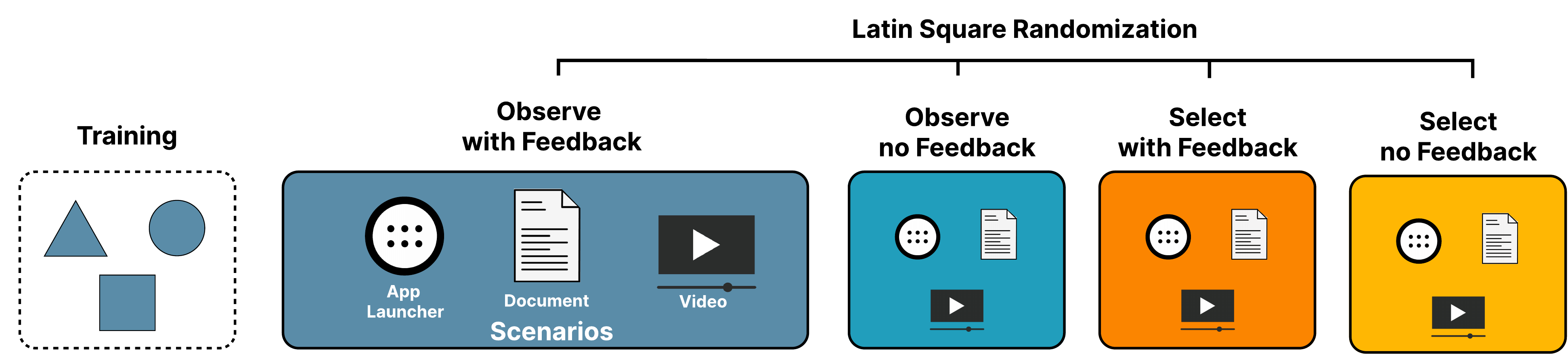}
    \caption{\textbf{Overview of the experimental Procedure.} Participants first completed a short training phase in which neutral geometric shapes (triangle, circle, square) were used instead of real icons. The main experiment followed, where each participant performed all four conditions (\textsc{Intent}: Observe / Select × \textsc{Feedback}: With / No) across three everyday MR scenarios (App Launcher, Document, Video). The order of blocks was counterbalanced using a balanced Latin Williams square design. Refer to \autoref{sec:procedure} for a complete description of the procedure and to \autoref{sec:task} for a complete descriptions of the scenarios in the task.}
    \Description{Screenshots of the three MR usage scenarios and training condition. Top row: (a) training interface with neutral geometric icons (triangle, circle, square), (b) document editor showing a “Hello World!” text window with toolbar buttons, (c) application launcher with icons for Files, Music, Settings, Safari, and TV, and (d) video player displaying a globe with playback controls. Bottom row: feedback screens, including gaze-click confirmation (“Gaze-Clicked Successfully!”), document export message, file browser with folders, and video playback interface.}
    \label{fig:procedure}
    \Description{Overview of the experimental design. On the left, participants completed a training phase using neutral geometric shapes (triangle, square, circle). The main task included three scenarios (App Launcher, Document, Video) under four experimental conditions: Observe with Feedback, Observe without Feedback, Select with Feedback, and Select without Feedback. Conditions were presented in counterbalanced order using Latin square randomization.}
\end{figure*}

The study began with a training phase in which participants familiarized themselves with the task. To prevent prior exposure to the experimental stimuli, we used neutral icons, i.e., circle, square, and triangle blue-shaped icons, that did not appear in the later scenarios. During training, participants performed five practice trials for each condition, ensuring they understood the required actions before entering the main experiment.

At the beginning of each trial, an instruction screen specified whether participants should \textit{select} or \textit{observe} the upcoming icon and whether feedback would be provided. The instruction also displayed the actual target icon or UI element for that trial, to ensure participants knew in advance which element to fixate. The instructions were as follows:  
\textit{(a) Select – No Feedback:} “Fixate on the target icon to select it. No feedback will be shown after your selection.”  
\textit{(b) Select – Feedback:} “Fixate on the target icon to select it. A short pop-up feedback will confirm your selection before the action continues.”  
\textit{(c) Observe – No Feedback:} “Fixate on the target icon to observe it,  keep your gaze on the icon. No feedback will be shown.”  
\textit{(d) Observe – Feedback:} “Fixate on the target icon to observe it, keep your gaze on the icon. A short pop-up feedback will appear after your fixation.”

\begin{figure*}[t]
    \centering
    \includegraphics[width=\linewidth]{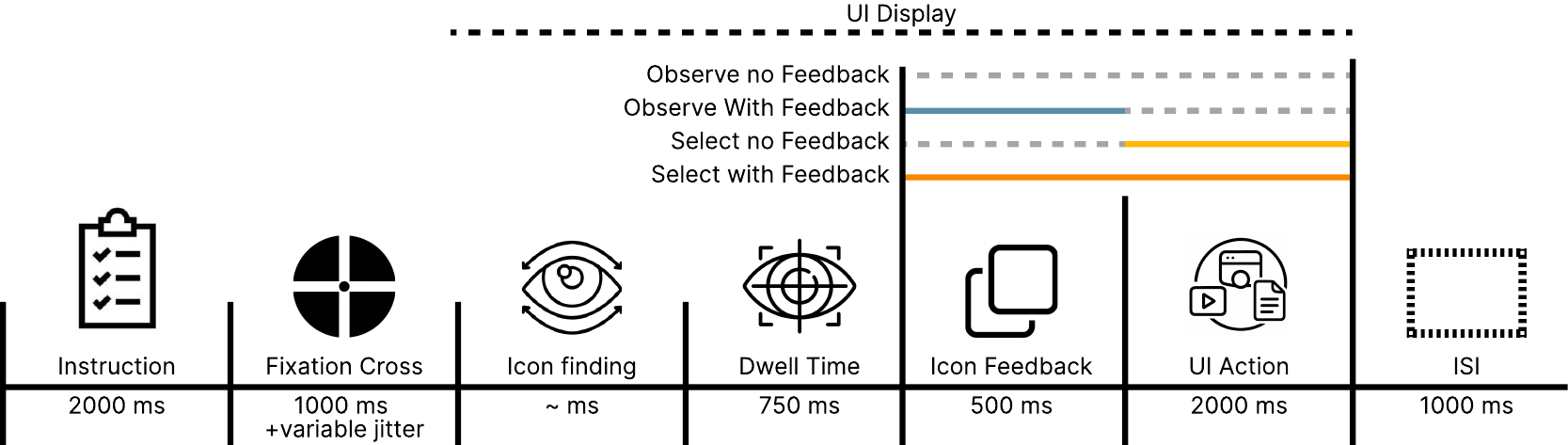}
    \caption{\textbf{Schematic of the four trial types (\textit{Select} vs.\ \textit{Observe}, with and without \textit{Feedback}).} 
    All trials began with Instruction (2000 ms), a Fixation Cross (1000 ms + variable jitter of 250, 500, or 750 ms), and a Dwell Time (750 ms). 
    The subsequent phases differed by condition. In the \textit{Select – No Feedback} condition {\textcolor[HTML]{FFB703}{\rule{0.25cm}{0.25cm}}}, 
    the dwell was followed directly by a UI Action (2000 ms) and then the ISI (1000 ms). 
    In the \textit{Select – Feedback} condition {\textcolor[HTML]{FB8500}{\rule{0.25cm}{0.25cm}}}, 
    the dwell was followed by a 500 ms Icon Feedback signal, then a UI Action (2000 ms) and the ISI (1000 ms). 
    In the \textit{Observe – No Feedback} condition {\textcolor[HTML]{219EBC}{\rule{0.25cm}{0.25cm}}}, 
    the dwell was followed immediately by the ISI (1000 ms) without any system response. 
    In the \textit{Observe – Feedback} condition {\textcolor[HTML]{5A8CA8}{\rule{0.25cm}{0.25cm}}}, 
    the dwell was followed by a 500 ms Icon Feedback signal and then the ISI (1000 ms). 
    Continuous lines indicate that a given phase (Icon Feedback or UI Action) was present in that condition, 
    while dotted lines indicate that the phase was absent. Each condition comprised 90 trials per participant. 
    In Observe trials, NO system action occurred. the interface remained static (No Feedback) or displayed only brief confirmatory feedback (With Feedback). 
    Only Select trials triggered consequential UI changes.}
    \label{fig:trial_structure}
    \Description{Timeline schematic showing all four trial types in a unified layout. Common phases (instruction, fixation cross with jitter, dwell time, and ISI) are shown in color across conditions. Optional phases are indicated with line-dotted boxes.}
\end{figure*}

The experimental phase was structured in blocks following a Latin Williams square design~\cite{wang2009construction}. Each block contained 90 trials and followed the trial structure illustrated in \autoref{fig:trial_structure}. The full study lasted approximately 75 minutes per participant, including setup, training, and breaks. We depict the procedure in \autoref{fig:procedure}.

\subsection{Trial Structure}

\begin{figure*}[htbp]
    \centering
    \includegraphics[width=\linewidth]{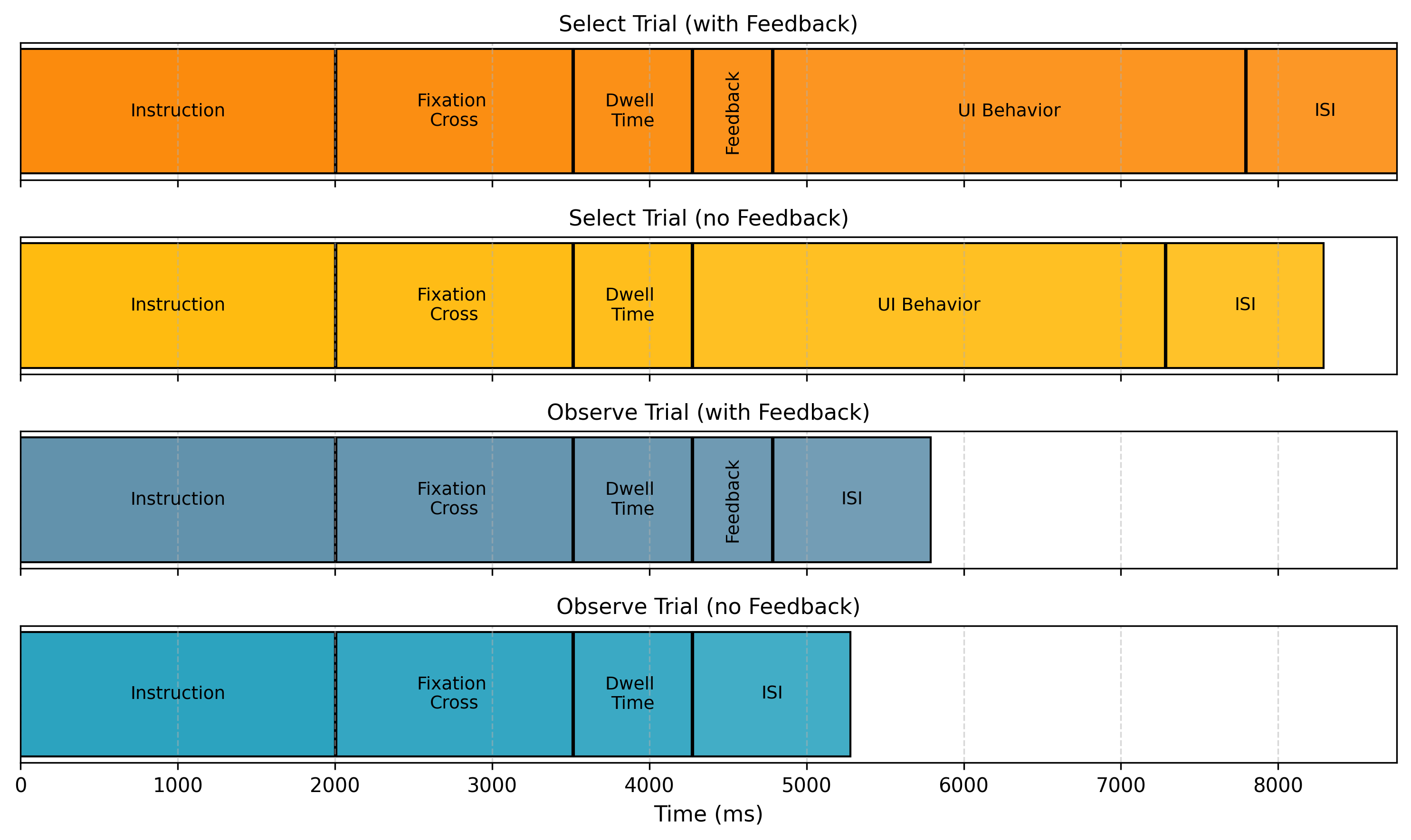}
    \caption{
       \textbf{Trial structure across Intent and Feedback conditions.}
The timeline illustrates the sequential phases of a single trial across four experimental conditions: Select (with and without feedback) and Observe (with and without feedback). 
All trials began with a 2-second instruction screen, followed by a fixation cross with a jittered duration (1000 + 250 / 500 / 750~ms). Participants then fixated on a UI element for 750~ms to select or observe it. 
In Select trials, a UI interaction was triggered for 3000~ms, coherent with the intention to Select. In Observe trials, no interaction followed the gaze dwell. When feedback was enabled, it was presented immediately after dwell time for 500~ms, preceding the UI phase. 
Finally, all trials concluded with a 1-second inter-stimulus interval (ISI) to allow attentional reset. 
For visualization purposes, the figure aligns all conditions to the same trial duration (8500~ms).
}
    \label{fig:trial_timeline}

    \Description{Timeline comparison of trial structures across the four experimental conditions: Select with Feedback, Select without Feedback, Observe with Feedback, and Observe without Feedback. Each condition begins with an instruction screen and a fixation cross, followed by a dwell time. In Select trials, feedback is shown immediately after the dwell for 500 ms and then followed by a UI action, whereas in the no-feedback condition the UI action follows directly. In Observe trials, no UI action occurs; feedback is only shown in the feedback condition immediately after the dwell. All trials conclude with an inter-stimulus interval (ISI). Timelines are displayed with a common total duration for ease of comparison.}
\end{figure*}

Our trial structure was inspired by previous work~\cite{reddy2024towards, chiossi2024searching}, following a real-world MR interaction paradigm, as illustrated in \autoref{fig:trial_structure}. The structure of the task, consisted of the following sequence: (1) participants were presented with a task description lasting two seconds to instruct which app to either \textit{Select} or \textit{Observe}; (2) a fixation cross (“\(+\)”) appeared at the center of the camera rig, based on \citet{thaler2013fixation}, which participants were required to fixate for a pseudorandom duration (1250, 1500, or 1750~ms) \cite{chiossi2024understanding, chiossi2024searching}; (3) following successful fixation, a UI element was presented that participants either observed or selected, requiring a dwell time of 750~ms~\cite{reddy2024towards}; (4) in \textit{Select} trials, this dwell triggered a UI interaction phase lasting 3000~ms. In trials with feedback, an additional feedback screen was presented for 250~ms after the interaction (or dwell) phase; (5) the trial concluded with a 1-second inter-stimulus interval (ISI), during which a blank screen was shown to facilitate attentional reset and reduce cognitive carryover effects~\cite{atkinson1989and, woodman1999electrophysiological}. For a graphical depiction of the trial timeline refeer to \autoref{fig:trial_timeline}.

Based on this structure, the total expected trial durations varied across conditions: in the \textit{Select with Feedback} condition, the trial lasted  8500~ms; in the \textit{Select without Feedback} condition, 8250~ms; in the \textit{Observe with Feedback} condition, 5500~ms; and in the \textit{Observe without Feedback} condition, 5250~ms. These differences reflect the presence or absence of the UI interaction and feedback phases, which are specific to the task condition.

\subsection{Stimuli}

In our study, we used real app icons while controlling for color to ensure consistency and ecological validity.
Recent findings by Liu et al.~\cite{liu2021app} emphasize the importance of icon design in optimizing search efficiency and user experience. Their results show that icons with varied colors significantly improve search efficiency by reducing cognitive load, fixation duration, and task completion times compared to uniform-colored icons. Furthermore, rounded square (RS) icons outperform circular or mixed-shape icons, yielding faster identification, fewer eye fixations, and higher user satisfaction.  The interaction between color and shape highlights that combining rounded square icons with varied colors achieves the best performance, enhancing search speed and minimizing cognitive effort.

Beyond icon design, perceptual factors such as object size and viewing distance impact interaction performance in MR. 
We therefore adopted icon sizes comparable to prior MR studies, where targets subtended visual angles of $8^{\circ}$–$20^{\circ}$~\cite{reddy2024towards}. 
This choice is consistent with usability guidelines showing that larger buttons ( $\approx$ $3^{\circ}50'$) improve speed and satisfaction compared to smaller ones ($\approx$ $1^{\circ}55'$)~\cite{hussain2023effects}. 
Moreover, a near-interaction distance of around 80~cm, close to average arm reach, has been recommended to maximize comfort and efficiency \cite{poston_human_2000, rasch2025ar}. 
Together, these findings support our use of icons subtending $\sim$3.5$^{\circ}$ at $\sim$80~cm as an optimal balance between usability, speed, and perceptual accessibility in MR.

\subsubsection{ERP Trials Amount Rationale}

To ensure adequate statistical power and waveform reliability, we based our trial count on empirical simulations and guidelines specific to ERP research  \cite{clayson2019methodological}. While participant sample size primarily determines power in repeated-measures designs, increasing the number of trials per condition enhances signal-to-noise ratio and improves measurement precision, which is particularly relevant for late ERP components with high intra-individual variability \cite{boudewyn2018many, clayson2019methodological}. \citet{jensen2022towards} recommends aiming for 80–150 trials per condition in within-subject ERP paradigms when targeting medium-sized effects with moderate trial-level noise. Accordingly, we presented 90 trials per condition, balancing practical constraints to detect condition-related differences in SPN amplitude with sufficient power and waveform stability.

\subsection{Apparatus}

We developed the virtual environment using Unity (Long-Term Support version 2022.3.x) and presented it through a Varjo XR-4 MR headset (Varjo, Finland, 200 Hz). 
The XR-4 provides dual mini-LED displays with a resolution of $3840 \times 3744$ pixels per eye ($\approx 51$ pixels per degree) and a $120^{\circ} \times 105^{\circ}$ field of view.
For environment tracking, we employed three SteamVR 2.0 base stations, following the manufacturer’s recommendations. The XR-4 headset was tethered to a Windows 11 workstation (HP Z1 Entry Tower G6) equipped with an Intel Core i9 processor running at 3.8\,GHz and 32\,GB of RAM.

\subsubsection{EEG Recording \& Preprocessing}
\label{sec:eeg_preprocessing}
We acquired EEG data (sampling rate = 500 Hz) via LiveAmp amplifier (Brainproducts, Germany) from 64 water-based electrodes from the R-Net elastic cap (Fp1, Fz, F3, F7, F9, FC5, FC1, C3, T7, CP5, CP1, Pz, P3, P7, P9, O1, Oz, O2, P10, P8, P4, CP2, CP6, T8, C4, Cz, FC2, FC6, F10, F8, F4, Fp2, AF7, AF3, AFz, F1, F5, FT7, FC3, C1, C5, TP7, CP3, P1, P5, PO7, PO3,
Iz, POz, PO4, PO8, P6, P2, CPz, CP4, TP8, C6, C2, FC4, FT8,
F6, F2, AF4, AF8). We kept impedance levels below $\le$20 k$\Omega$. We set the reference at FCz during the recording, while FPz was used as ground. We placed the electrodes using the International 10-10 layout. For time synchronization with the MR environment, we employed the Lab Streaming Layer Framework, while for preprocessing and analysis, we used the MNE-Python Toolbox~\cite{gramfort2013meg}. We first automatically detected bad or outlier channels via the random sample consensus (RANSAC) method~\cite{bigdely2015prep} of spherical splines for estimating scalp potential based on algorithms proposed by Perrin~\cite{perrin1989spherical}.  We then applied a notch filter (50 Hz) and re-referenced to the common average reference (CAR). We then band-passed the signal between (1-15 Hz) to remove high and low-frequency noise. We applied an Independent Component Analysis (ICA) for artifact detection and correction with extended Infomax algorithm~\cite{lee1999independent}. We automated the labeling and rejection process of ICA components via the MNE plugin ``ICLabel ''~\cite {li2022mne}.  We rejected epochs that showed blinks, eye movement, muscle, or single-channel artifacts in any of the electrodes using Bayesian optimization for threshold selection \cite{jas2017autoreject}. On average, we removed $6.07$ ($SD=9.89$) independent components on average for each participant.

\subsubsection{ERP Analysis}
We segmented the continuous signal between -1000 ms and 0 ms before UI Display. ERPs were baseline corrected from -1000 to -750 ms as we were primarily interested in the -750 to 0ms window where an SPN is to be expected. We avoided the -850 to -750ms range due to the fixation-related Lambda response resulting from the first fixation on the target stimulus \cite{ries2018Lambda}. The SPN was thus quantified as negative peak amplitudes in the -750 ms -- 0 ms range \cite{reddy2024towards}. For SPN computation, we chose electrodes O1, Oz, O2, Iz, PO7, PO3, POz, PO4, PO8, P5, P6, P7, P8, P9, and P10 based on previous work~\cite{reddy2024towards}.

\subsection{Statistical Modelling}
We compared a set of linear mixed models predicting SPN amplitude (\textit{neg\_peak}) from \textit{Intention} (Observe vs.\ Select) and \textit{Feedback} (Feedback vs.\ No Feedback). Random-effect structures were incrementally expanded from including only participant (PID) to adding \textit{Order}, \textit{channel}, and \textit{Scene} as grouping factors. Model comparison was performed using maximum likelihood estimation and likelihood ratio tests \cite{vrieze2012aicbic}. The best-fitting model, selected with the lowest Akaike Information Criterion (AIC), included random intercepts for PID, channel, and trial. Full model fitting results are reported in the Appendix in \autoref{app:table}.

To assess generalizability across interaction contexts, we conducted a sensitivity analysis with \textit{Scene} as a fixed factor. We compared models with \textit{Scene} as: (1) a random effect (baseline), (2) a fixed main effect, (3) with two-way interactions (\textit{Intention:Scene}, \textit{Feedback:Scene}), and (4) with the three-way \textit{Intention} $\times$ \textit{Feedback} $\times$ \textit{Scene} interaction. This tests whether the core pattern replicates across the three scenarios or requires specification of boundary conditions.

\subsection{Sample Size Justification \& Participants}

An a priori power analysis was conducted using G*Power (version 3.1), to estimate the required sample size for a study analyzed with linear mixed-effects models \cite{faul2009statistical}. The analysis assumed a medium effect size ($f = .25$) based on HCI guidelines \cite{yatani2016effect}, an alpha level of $.05$, and a desired power of $.80$. The correlation among repeated measures was set to $.50$, and the nonsphericity correction was set to $1.00$. The results indicated that a minimum total sample size of $24$ participants is required to achieve a desired statistical power (actual power = $.817$).
A total of $28$ participants took part in the study, resulting in an actual power of $.88$. The sample included $20$ males and $8$ females, none diverse. Participants' ages ranged from $18$ to $32$ years (mean = $23.6$, SD = $3.1$). The participants’ familiarity with AR, AV, and VR technologies was assessed, following previous work \cite{chiossi2024searching, chiossi2024understanding}. All participants had prior exposure to AR (M = $3.2$, SD = $1.4$), AV (M = $2.9$, SD = $1.6$), and VR (M = $4.1$, SD = $1.7$) technologies, rated on a 7-point scale ranging from 1 (not familiar at all) to 7 (extremely familiar). Exclusion criteria for participants included a medical history of psychological or neurological disorders, color blindness, and visual impairments.

\subsection{Classification}
\label{sec:classification}
We evaluated two complementary decoding strategies to assess the robustness of
multimodal EEG-based intention classification: a \textit{person-dependent}
setting, which estimates how well models adapt to individual neural patterns,
and a \textit{person-independent} setting, which evaluates generalization to
unseen users.

\subsubsection{Data Preparation}
For the person-dependent models, a separate dataset was created for each
participant. Trials were split into training (60\%), validation (20\%), and
test (20\%) subsets on a per-participant basis. An additional 80/20 split of the training data was used during hyperparameter optimization to prevent data leakage and ensure that the final test set remained unseen.

For the person-independent models, all trials from all participants were first aggregated and assigned participant identifiers. The dataset was then
partitioned \textit{at the participant level}: 60\% of participants were used for training, 30\% for validation, and the remaining 10\% for testing. This ensured that no trial from any test participant influenced model training or hyperparameter tuning.

\subsubsection{Feature Extraction}
For both settings, we used the same preprocessing pipeline described in
\autoref{sec:eeg_preprocessing}. Epochs were extracted from the SPN analysis window (–750 to 0~ms) and baseline-corrected to the –1000 to –750~ms interval.
Trials were represented as $(N, C, T)$ tensors and passed directly to the CNN architectures, allowing models to learn hierarchical spatiotemporal features of anticipatory EEG activity without manual feature engineering
\citep{lotte2018review, craik2019deep}.

\subsubsection{Model Selection}
We evaluated five established deep learning architectures for EEG decoding:
\textit{EEGNetv4} \citep{lawhern2018eegnet}, \textit{ShallowFBCSPNet} and
\textit{Deep4Net} \citep{schirrmeister2017deep}, \textit{EEGResNet}
\cite{chen2025steady}, \textit{EEGInceptionERP} \cite{santamaria2020inception}. Models were trained with 100~epochs in the person dependent analyses and 100~epochs in the person independent setting, using Adam, AdamW, or RMSProp optimizers (sampled during hyperparameter search). Early stopping on validation accuracy mitigated overfitting.

\subsubsection{Hyperparameter Optimization}
Hyperparameter optimization was performed using the \texttt{Optuna} framework \cite{akiba2019optuna}. In the person-dependent setting, optimization was conducted separately for each participant using the inner validation split.
In the person independent setting, hyperparameters were tuned exclusively on trials from the validation participants, with 540 trials per architecture.
Search spaces included filter sizes, kernel lengths, depth multipliers,
pooling parameters, activation functions, batch sizes, and optimizer settings. Each model was subsequently retrained using the best-performing configuration.

\subsubsection{Evaluation}
Evaluation procedures matched the goals of each setting. For the
person-dependent models, performance was computed on each participant’s
held-out test data and compared across architectures using paired-samples
\textit{t}-tests on participant-wise accuracies. For the person-independent
models, performance was evaluated on trials from test participants only,
representing true zero-overlap generalization to unseen individuals. Because this setting yields a single pooled value per model, we report accuracy differences and bootstrap confidence intervals (10{,}000 resamples) \cite{tibshirani1993introduction, combrisson2015exceeding} instead of participant-wise paired statistics.

\paragraph{LIME-Based Interpretability Analysis}

To examine how our neural networks relied on specific EEG features, we employed Local Interpretable Model-agnostic Explanations (LIME), a perturbation-based, model-agnostic framework that explains individual predictions through locally linear surrogate models \cite{ribeiro2016lime}. For each EEG trial, LIME generates a set of perturbed variants by randomly masking spatiotemporal segments and evaluating how these perturbations alter the network’s output. A weighted linear regression is then fitted over these perturbation–response pairs, with higher weights assigned to perturbed samples that remain closer to the original input. The resulting coefficients form a saliency map that highlights the EEG channels and time points most influential for that specific prediction. Recent work demonstrates that LIME is effective in capturing stable and physiologically meaningful explanatory patterns in deep EEG models \cite{rejer2025averagedlime}, making it well-suited for interpreting our classifier’s reliance on preparatory/SPN-related neural activity.

\section{Results}
We present results from both ERP analyses of the SPN component and deep learning–based classification of user intention. For a summary of the results, we refer the reader to \autoref{sec:summary_results}.

\subsection{SPN Amplitude}



\begin{figure*}[htbp]
    \centering
    \includegraphics[width=\textwidth]{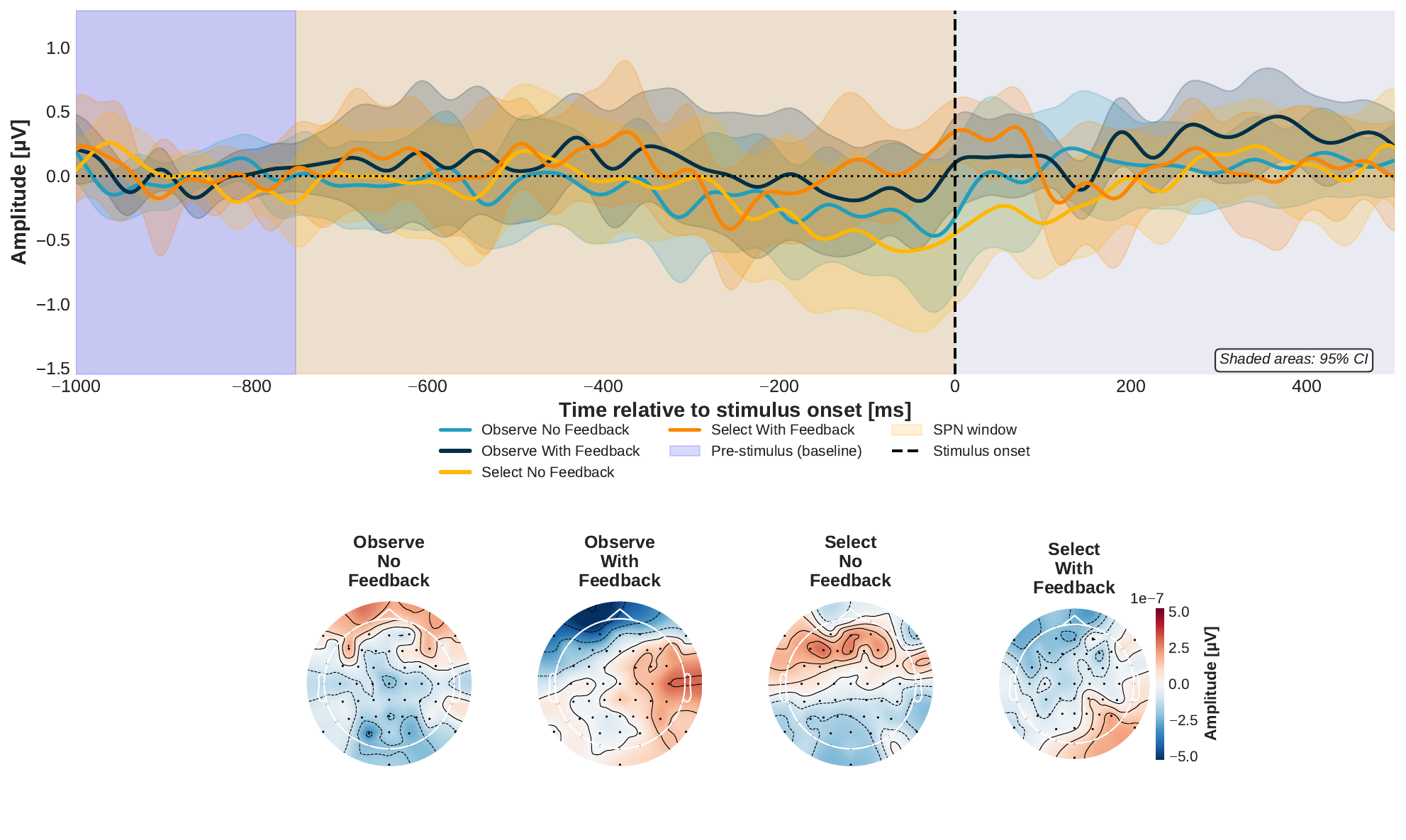}
    \caption{
    \textbf{Grand-average ERP waveforms and SPN scalp topographies across experimental conditions.} 
    The ERP plot displays the grand-average activity at the posterior ROI ( O1, Oz, O2, Iz, PO7, PO3, POz, PO4, PO8, P5, P6, P7, P8, P9, and P10) with shaded areas indicating the 95\% confidence intervals. 
    The pre-stimulus baseline window (–1000 to –750~ms) is shown in purple, and the SPN window (–750 to 0~ms) is shaded in beige; the vertical dashed line indicates stimulus onset (0~ms). 
    The scalp maps below illustrate mean voltage distributions during the SPN interval for each condition. 
    Observe--No Feedback elicited the strongest anticipatory negativity, whereas both the intention to select and the presence of feedback reduced the SPN magnitude.
    }
    \label{fig:erp_spn_multi_panel}
    \Description{
    Grand-average ERP waveforms are shown for four conditions, with shaded 95\% CIs and marked SPN and baseline windows. 
    Scalp topographies below the waveform display spatial distributions of the SPN for each condition. 
    Observe without feedback shows a stronger right-posterior negativity compared to all other conditions.}
\end{figure*}

\begin{figure*}[htbp]
    \centering
    \includegraphics[width=\textwidth]{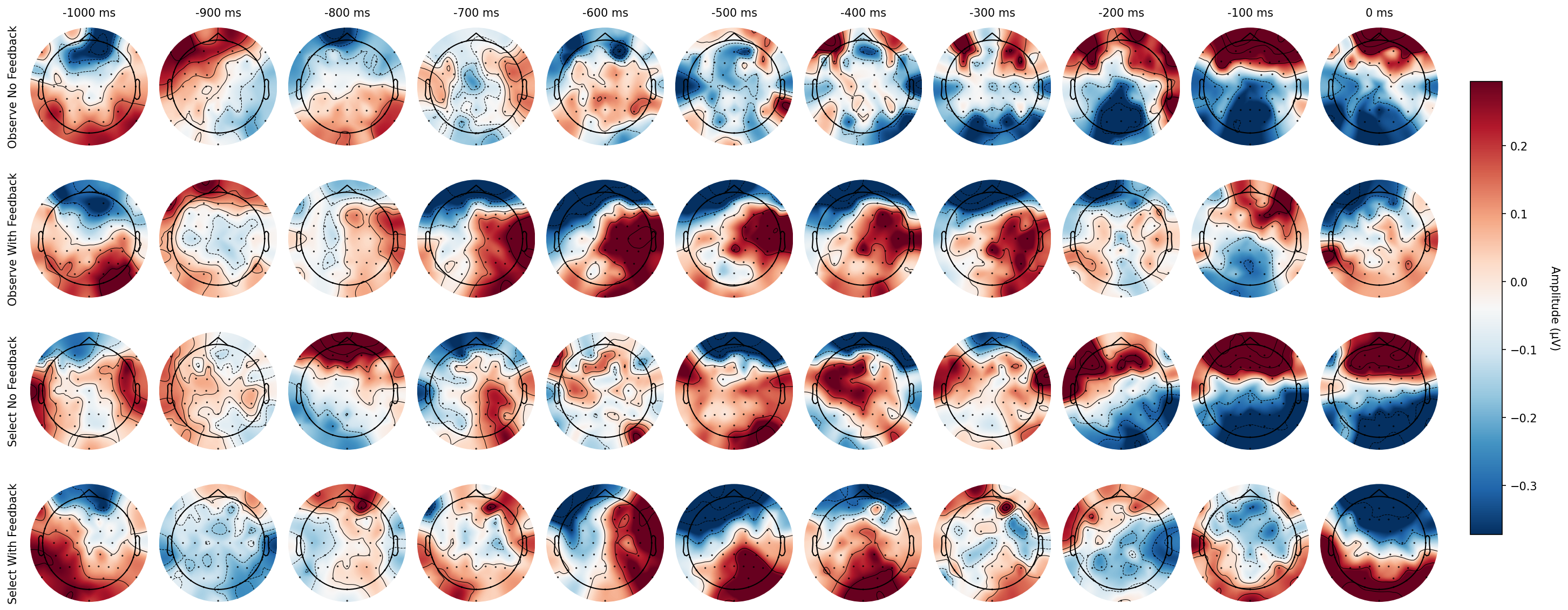}
    \caption{ \textbf{Topographic maps of ERP activity from –1000~ms to stimulus onset (0~ms) across the four experimental conditions.} 
Each column represents a 100~ms step, and rows correspond to conditions (Observe No Feedback, Observe With Feedback, Select No Feedback, Select With Feedback). 
A sustained posterior negativity (blue shading) is evident over occipito-parietal electrodes during the anticipatory interval, 
with the strongest SPN emerging in the Observe–No Feedback condition. 
Both the intention to select and the availability of feedback increased this negativity.
}
\label{fig:topomap}
\Description{Topographic maps of ERP activity from –1000 ms to stimulus onset (0 ms) for the four experimental conditions: Observe without Feedback, Observe with Feedback, Select without Feedback, and Select with Feedback. Each column shows 100 ms time steps, with blue indicating negative voltage and red indicating positive voltage. A sustained posterior negativity appears over occipito-parietal electrodes, strongest in the Observe without Feedback condition.}
\end{figure*}

We fitted a linear mixed model, estimated using restricted maximum likelihood (REML) and the \texttt{nloptwrap} optimizer, to predict SPN amplitude (\textit{neg\_peak}) from \textit{Intention} (Observe vs.\ Select) and \textit{Feedback} (With vs.\ No). The model included random intercepts for participant (PID), order, channel, and scene. The model's explanatory power was substantial, with a conditional $R^2 = .33$, although the variance explained by the fixed effects alone was small (marginal $R^2 = .0013$). 

The model intercept, corresponding to Select trials with Feedback, was significantly negative, $\beta = -8.17$, 95\% CI [$-10.35$, $-6.00$], $t(71{,}901) = -7.36$, $p < .001$. Compared to this baseline, Observe trials showed significantly larger (less negative) amplitudes, $\beta = 1.02$, 95\% CI [0.85, 1.20], $t(71{,}901) = 11.45$, $p < .001$, standardized $\beta = .11$, 95\% CI [.10, .13]. Likewise, trials without Feedback were associated with significantly larger amplitudes, $\beta = .44$, 95\% CI [0.26, 0.61], $t(71{,}901) = 4.96$, $p < .001$, standardized $\beta = .05$, 95\% CI [.03, .07]. Importantly, the Intention $\times$ Feedback interaction was significant, indicating that the difference between Observe and Select trials was reduced in the absence of Feedback, $\beta = -0.96$, 95\% CI [$-1.21$, $-0.71$], $t(71{,}901) = -7.60$, $p < .001$, standardized $\beta = -.11$, 95\% CI [$-.14$, $-.08$].

\subsubsection{Sensitivity Analysis}

We assessed whether the \textit{Intention} $\times$ \textit{Feedback} pattern generalized across the three MR scenarios. Model comparison showed that two-way interactions with \textit{Scene} significantly improved fit ($\chi^2$(4) = 70.56, \textit{p} < .001, AIC = 508,389 vs. baseline AIC = 508,457). However, the three-way interaction did not further improve the model ($\chi^2$(2) = 1.64, \textit{p} = .440), indicating that the core pattern replicated consistently.

The \textit{Intention} $\times$ \textit{Feedback} interaction showed consistent effect sizes across scenarios: $\beta$ = $-0.88$ $\mu$V (SE = 0.13, \textit{z} = $-6.92$, \textit{p} < .001) for App Launcher, Document, and Video contexts. In all three scenarios, observation without feedback produced the strongest SPN, while selection intent and feedback presence both reduced anticipatory activity. This replication demonstrates that SPN reflects anticipatory uncertainty across diverse MR interaction contexts.


\subsection{Classification Results}

\subsubsection{Person-Dependent Classification}

Across participants, mean validation classification accuracies (\textit{M} $\pm$ \textit{SD}) were as follows: \textit{EEGInceptionERP} (\textit{M} = .784, \textit{SD} = .090, 95\% CI [.749, .819]), \textit{EEGResNet} (\textit{M} = .758, \textit{SD} = .080, 95\% CI [.727, .789]), \textit{ShallowFBCSPNet} (\textit{M} = .742, \textit{SD} = .080, 95\% CI [.711, .773]), \textit{EEGNetv4} (\textit{M} = .736, \textit{SD} = .067, 95\% CI [.710, .761]), and \textit{Deep4Net} (\textit{M} = .732, \textit{SD} = .078, 95\% CI [.702, .762]). Paired-samples \textit{t}-tests were conducted to compare model performance across participants. \textit{EEGInceptionERP} significantly outperformed \textit{EEGNetv4}, $t$(27) = -3.18, $p$ = .003; \textit{Deep4Net}, $t$(27) = -3.47, $p$ = .002; and \textit{ShallowFBCSPNet}, $t$(27) = 2.93, $p$ = .007. Additionally, \textit{EEGResNet} significantly outperformed \textit{Deep4Net}, $t$(27) = 2.48, $p$ = .020.

No significant differences were found between \textit{EEGNetv4} and \textit{EEGResNet}, $t$(27) = -1.84, $p$ = .077; \textit{EEGNetv4} and \textit{Deep4Net}, $t$(27) = 0.35, $p$ = .726; \textit{EEGNetv4} and \textit{ShallowFBCSPNet}, $t$(27) = -0.49, $p$ = .626; \textit{EEGResNet} and \textit{EEGInceptionERP}, $t$(27) = -1.71, $p$ = .099; \textit{EEGResNet} and \textit{ShallowFBCSPNet}, $t$(27) = 1.19, $p$ = .243; and \textit{Deep4Net} and \textit{ShallowFBCSPNet}, $t$(27) = -0.64, $p$ = .530.

These results indicate that \textit{EEGInceptionERP} provided the most reliable classification of user intention on a person-dependent basis. It significantly outperformed three of the four other models and demonstrated robust performance across participants. \textit{EEGResNet} also showed consistently high accuracy and significantly outperformed \textit{Deep4Net}.

\paragraph{Computational Footprint and Latency.}
Across all trained architectures, the computational requirements of the decoding pipeline were low in absolute terms. Model sizes ranged from 0.44\,MB (EEGNetv4, 115k parameters) to 4.87\,MB (Deep4Net, 1.27M parameters), with EEGInceptionERP positioned in between at 1.22\,MB (318k parameters). Inference times followed the same trend: EEGNetv4 required an average of 0.67\,ms (SD=0.34; median=0.59\,ms; p95=1.12\,ms; p99=1.25\,ms), EEGInceptionERP 1.75\,ms (SD=0.40; median=1.70\,ms; p95=2.38\,ms; p99=2.62\,ms), and Deep4Net 3.07\,ms (SD=0.91; median=2.85\,ms; p95=4.52\,ms; p99=4.89\,ms). All architectures achieved more than 7.6 predictions per second. These values were derived using the same SPN-based input representation (–750 to 0\,ms window) for both person-dependent and person-independent models.

To contextualize these values within an interactive pipeline, a full SPN window corresponds to 750\,ms of EEG acquisition, to which preprocessing and inference contribute less than 5\,ms in total. This latency is comparable to standard dwell-time selection in MR interfaces, 
which typically require 500-1000ms fixation durations \cite{pfeuffer2017gaze, 
majaranta2012communication}. The ~770-785ms window falls within this range, making 
SPN-based intention decoding feasible for adaptive systems that adjust 
dwell-time thresholds, provide graduated confirmations, or trigger contextual 
assistance. Importantly, the computational overhead (<5ms) is negligible 
relative to the inherent SPN acquisition window (750ms), meaning faster EEG 
hardware or reduced temporal windows could further decrease total latency 
without algorithmic bottlenecks. Real-time implementation would require 
streaming EEG preprocessing and online classification, which prior work has 
demonstrated for similar BCIs \cite{lotte2018review, ang2012clinical,lawhern2018eegnet}.

\subsubsection{Person Independent Classification}

For the person independent setting, each model yields a single pooled
validation accuracy because training and evaluation occur on disjoint
participant groups. 

To quantify uncertainty around the pooled estimates, we computed non-parametric bootstrap confidence intervals
(10{,}000 resamples) based on related work \cite{boudewyn2018many, jensen2022towards, combrisson2015exceeding}.

Mean accuracies with  95\% bootstrap confidence intervals were:
\textit{Deep4Net} ($M = .692$, 95\% CI [.673, .712]),
\textit{EEGNetv4} ($M = .662$, 95\% CI [.642, .681]),
\textit{ShallowFBCSPNet} ($M = .642$, 95\% CI [.621, .662]),
\textit{EEGInceptionERP} ($M = .631$, 95\% CI [.610, .651]),
and \textit{EEGConformer} ($M = .506$, 95\% CI [.487, .525]).

\subsubsection{LIME-Based Interpretability}

To examine which EEG features drove each classifier’s decisions, we performed LIME perturbation  analyses on correctly classified trials and aggregated saliency maps per participant and per model. 
Across architectures, feature importance was strongly localized over centro-parietal and  parieto–occipital sites corresponding to canonical slow negative SPN activity. 
Below we report model-specific topographies and quantitative interpretability metrics.

\paragraph{EEGNetv4.}
LIME maps for EEGNetv4 showed a focal cluster centered on \textit{Pz, CPz, Cz}, extending into 
\textit{POz, PO3, PO4, O1, O2, Iz}. Spatial and temporal concentration were near ceiling 
($H_\mathrm{spatial}=.99$, $H_\mathrm{temporal}=.98$), and inter-trial variability was minimal  ($SD<.001$), indicating stable and physiologically consistent reliance on posterior preparatory  components, see \autoref{fig:lime_eegnetv4}.

\paragraph{ShallowFBCSPNet.}
ShallowFBCSPNet produced a highly similar posterior pattern, with peak importance over \textit{CP1--CP4, P1--P4, P5/P6, PO3/PO4}. Concentration metrics remained extremely high  ($H_\mathrm{spatial}=.99$, $H_\mathrm{temporal}=.99$), with very low inter-trial dispersion  ($SD<.001$). The model emphasized midline and adjacent parietal channels consistent with  prior SPN literature, see \autoref{fig:lime_shallow}.

\paragraph{Deep4Net.}
Deep4Net saliency maps showed the broadest but still physiologically coherent distribution, centered  on \textit{Pz, CPz, P3/P4, P5/P6}, with contributions from \textit{O1/O2} and \textit{POz}. 
Concentration values remained high ($H_\mathrm{spatial}=.98$, $H_\mathrm{temporal}=.98$), and  inter-trial stability was again excellent ($SD<.001$), demonstrating consistent posterior  weighting across samples, see \autoref{fig:lime_deep4}.

\paragraph{EEGResNet.}
EEGResNet exhibited the most sharply localized pattern, with maximal saliency at the midline parietal cluster \textit{CPz--Pz--POz} and secondary contributions at \textit{P1/P2} and  \textit{PO3/PO4}. Both spatial and temporal concentration approached unity ($H_\mathrm{spatial}=.99$, $H_\mathrm{temporal}=.98$), and inter-trial variability was negligible  ($SD<.001$), indicating strongly convergent usage of SPN-related features, see \autoref{fig:lime_resnet}.

\paragraph{EEGInceptionERP.}
The Inception-based model showed a characteristic parietal--occipital ring, with peak importance at \textit{PO7/PO8, PO3/PO4, P5/P6, O1/O2, Iz}. As with the other architectures, entropy-based  concentration was extremely high ($H>.99$), and explanation stability was strong ($SD<.001$). 
The model’s weighting of late SPN timepoints was consistent with its high temporal filter resolution, see \autoref{fig:lime_inception}.


\section{Discussion}
This work provides the first demonstration that anticipatory EEG signals during realistic MR interaction not only reveal how users monitor system behavior under uncertainty, but also contain sufficiently discriminative information to support intention decoding through deep learning. Below, we summarize our main empirical findings before discussing our hypotheses, research questions and their implications for MR interaction design and adaptive interfaces.

\subsection{Summary of Results}
\label{sec:summary_results}
We examined how user intention (Select vs. Observe) and system feedback (With vs. Without) shape anticipatory neural activity during gaze-based interaction in MR. Participants interacted with familiar interface elements across everyday MR scenarios while EEG was recorded to measure the SPN.

Across all conditions, a robust SPN emerged, showing that anticipatory neural activity reliably manifests during naturalistic MR interaction and is not confined to controlled laboratory tasks. The SPN was strongest when participants merely observed an icon without receiving feedback, suggesting that SPN reflects heightened monitoring under uncertainty rather than intention preparation alone. Intention and feedback further interacted: in Select trials, SPN amplitudes remained stable regardless of feedback, whereas in Observe trials, feedback markedly reduced anticipatory activity. When both intention and feedback were present, SPN amplitudes again became more negative, indicating greater engagement when system responses aligned with user goals.

Sensitivity analysis confirmed that these patterns replicated across all three MR scenarios (App Launcher, Document, Video). While effect magnitudes varied by context, likely reflecting differences in task demands or visual complexity, the fundamental \textit{Intention} $\times$ \textit{Feedback} interaction remained consistent across all scenarios. This replication demonstrates that SPN's sensitivity to anticipatory uncertainty generalizes across diverse MR interaction types rather than being limited to specific interface contexts.

Beyond the ERP findings, we evaluated whether user intention could be decoded from anticipatory EEG. In the person-dependent setting, deep learning models classified Select vs.\ Observe trials with high accuracy (mean performance up to 78\%; individual participants up to 97\%), with \textit{EEGInceptionERP} showing the strongest performance. In the person independent setting, i.e., evaluating generalization to unseen users, accuracy decreased (best model: \textit{Deep4Net}, $M = .692$, 95\% CI [.673, .712]). Importantly, across both settings, the two classes were detected equally well: per-class accuracies differed by less than 3–5\%, and no architecture exhibited systematic bias toward either intention class. Across all architectures, LIME-based saliency analyses revealed highly consistent and physiologically plausible patterns, with every model relying on centro–parietal and parieto–occipital SPN components to distinguish Select from Observe trials. Despite architectural differences, the models converged on similar anticipatory neural features.

Taken together, these findings contribute in four complementary ways. First, they confirm that the SPN is robustly elicited across ecologically valid MR interactions, establishing it as a reliable marker of anticipatory processing outside traditional laboratory tasks. Second, they refine theoretical interpretations by showing that the SPN indexes anticipatory uncertainty rather than motor preparation, shaped by the joint influence of intention and system feedback. Third, they demonstrate that user intention can be decoded from anticipatory EEG using deep learning, with high reliability in a person-dependent setup and meaningful generalization in person-independent models. Finally, they highlight implications for adaptive MR design: SPN-derived signals could help mitigate false activations (i.e., the Midas Touch problem), guide the timing and richness of system feedback, and be integrated with other modalities, such as pupil size or electrodermal activity, to support more intention-aware and uncertainty-sensitive interfaces.

\subsection{H1: SPN is a Robust Marker of Anticipation in MR}

Our first hypothesis predicted that SPN would be detectable during gaze-based MR 
interactions across all task conditions. This was supported: we observed reliable 
SPN activity in every experimental condition and scenario, demonstrating that 
anticipatory signals can be measured consistently in MR.

This finding extends SPN research beyond highly constrained laboratory tasks to 
ecologically valid MR contexts where users interacted with everyday UI elements. 
Earlier studies reported SPN-like negativity in constrained settings such as 
gaze-controlled games~\cite{shishkin2016eeg} or VR selection~\cite{reddy2024towards}. 
Our results extend this evidence to everyday MR tasks, including app launching, 
document editing, and media playback, demonstrating that SPN is not limited to 
simplified paradigms but also emerges in more ecologically valid interaction contexts.

Critically, our design isolated anticipatory processes from motor preparation by 
eliminating overt motor actions, as all interactions relied solely on gaze, with no button presses or hand movements that would introduce motor-related potentials 
such as the RP or LRP 
\cite{shibasaki2006components, eimer1998lateralized}. This exclusion of motor 
components allows us to attribute the observed negativity specifically to 
anticipatory monitoring rather than action preparation. The centro-parietal scalp 
distribution we observed (see \autoref{fig:topomap}) aligns with canonical SPN topography reported 
in non-motor anticipation tasks \cite{vanBoxtel2004cortical, brunia2012negativeslow}, 
further supporting the interpretation that these signals reflect cognitive 
anticipation of upcoming events rather than motor readiness. Moreover, the pattern 
that Observe trials, particularly those without feedback, elicited stronger SPN than 
Select trials directly contradicts what motor preparation accounts would predict, 
since committing to act should increase motor-related activity rather than decrease it. 
Instead, our findings align with theoretical accounts positioning SPN as a marker 
of anticipatory uncertainty and outcome monitoring \cite{moris2013learning, 
walentowska2018relevance, seidel2014uncertainty}, where the brain allocates 
preparatory resources in proportion to the unpredictability of forthcoming events.

\subsection{H2: Observation Elicits Stronger Anticipatory Activity Than Selection}

Our second hypothesis predicted stronger SPN amplitudes during Intent-to-Select trials, assuming that goal-directed actions require greater anticipatory engagement. Results pointed in the opposite direction: Observe trials, especially those without feedback, produced the most pronounced SPN.

This unexpected finding challenges traditional interpretations of SPN as primarily preparatory \cite{brunia2012negativeslow} instead supports accounts linking it to anticipatory uncertainty and monitoring demands~\cite{moris2013learning, walentowska2018relevance}. Passive observation without feedback created outcome uncertainty, producing heightened anticipatory activity, while committing to selection reduced ambiguity about upcoming events, leading to weaker SPN amplitudes. Earlier work in VR and gaze-controlled tasks~\cite{reddy2024towards, shishkin2016eeg} reported larger SPN during active selection, but in those designs intention was inseparable from a guaranteed system response. Our factorial approach separates these factors, showing that SPN is more sensitive to uncertainty about outcomes than to the intention to act itself. 

This theoretical refinement repositions SPN as a marker of expectancy and uncertainty management, modulated by outcome predictability rather than action preparation alone.

\subsection{H3: Intention and Feedback Jointly Shape Anticipation}

Our third hypothesis predicted that intention effects would remain consistent regardless of feedback presence, following \citet{reddy2024towards} findings that SPN is driven by intent rather than feedback. This hypothesis was not supported. Instead, we found a significant interaction between intention and feedback conditions, showing a pattern not previously described in the SPN literature.

In Select trials, SPN amplitudes stayed stable across feedback conditions, suggesting that once users commit to acting, anticipation is relatively insensitive to whether feedback is present. In Observe trials, however, feedback markedly reduced SPN, showing that passive monitoring is strongly shaped by outcome predictability.

This interaction represents a novel finding that SPN reflects both internal goals and external contingencies rather than being purely intention-driven. 
Classic and review work already argued for a non-motoric SPN sensitive to upcoming events~\cite{vanBoxtel2004cortical,brunia2012negativeslow,chwilla1991event}, and recent studies show that relevance and uncertainty combine to shape SPN during feedback anticipation~\cite{walentowska2018relevance}. In VR selection tasks, larger SPN during active intent has been reported~\cite{shishkin2016eeg,reddy2024towards}, yet those paradigms intertwined intention with guaranteed feedback. 

Our manipulation separates these factors and suggests that SPN tracks anticipatory monitoring under outcome uncertainty: feedback availability strongly attenuates SPN during passive observation, but has little impact once users commit to acting. Related evidence that SPN scales with uncertainty or attention allocation~\cite{seidel2014uncertainty, kotani2025stimulus} supports this refinement and helps explain why SPN can peak in Observe–No Feedback despite weaker action demands.

Together with prior work showing that SPN scales with relevance and uncertainty during outcome anticipation \cite{walentowska2018relevance,seidel2014uncertainty} and with XR studies linking intent-to-select to larger pre-event negativity \cite{shishkin2016eeg,reddy2024towards}, our results suggest that SPN reflects the balance between internal goals and outcome predictability rather than intention alone.

\subsection{RQ: Decoding Intention from EEG}

Our research question asked to what extent intention (\textsc{Select} vs.\ \textsc{Observe}) can be decoded from EEG signals within participants using deep learning. To address this, we conducted offline classification analyses with five established CNN architectures (EEGNetv4, ShallowFBCSPNet, Deep4Net, EEGResNet, and EEGInceptionERP) in a person-dependent setup, with hyperparameters tuned separately for each participant.

Classification accuracies ranged widely from 75\% to 97\% across participants, highlighting substantial individual differences that would require personalized calibration in practical applications. EEGInceptionERP achieved the highest mean accuracy (78.4\%), followed by EEGResNet (75.8\%), though the large variance indicates that some participants exhibited more separable neural patterns than others. 

These offline results demonstrate that anticipatory EEG activity during the -750 to 0ms window contains discriminative information about user intention. However, several limitations constrain the immediate applicability to real-time MR systems: the person-dependent approach necessitates individual training and the computational requirements and latency constraints of online classification remain untested. Even so, the results establish a computational basis for intention-aware MR systems, showing that user goals can be inferred from anticipatory EEG and pointing toward the next step of validating such methods in real-time, interactive MR environments.

\subsection{Implications for Adaptive MR Interfaces}

Our findings showed that SPN is consistently elicited in MR (\textbf{H1}), is shaped by both intention and feedback (\textbf{H2}–\textbf{H3}), and can be decoded with reasonable accuracy within participants (\textbf{RQ}). Rather than treating SPN as a simple binary marker of intent, these results point to its broader role as an indicator of an anticipatory state that integrates intention with uncertainty and monitoring demands.

Our results indicate that SPN reflects not only intention but also anticipatory uncertainty. This broader view opens several design directions for adaptive MR. Increased SPN during uncertain observation could prompt the system to offer lightweight cues or confirmations, while stable SPN during committed selections suggests opportunities to streamline interaction by reducing redundant steps. Variability in SPN patterns may further differentiate exploratory from committed gaze, informing adaptive dwell thresholds or graduated feedback strategies. Finally, person-dependent decoding demonstrates the feasibility of classifiers that adapt to user-specific uncertainty profiles. Combined with complementary signals such as pupil size and EDA, SPN could contribute to a multimodal pipeline for uncertainty detection. Pupil-linked arousal reliably encodes decision uncertainty and shapes choice behavior~\cite{fan2023pupil,he2024pupil}, while EDA captures anticipatory arousal in emotionally salient and uncertain contexts~\cite{wichary2016probabilistic, studer2016psychophysiological}. Together, these measures provide converging evidence of user uncertainty, enabling more robust disambiguation of gaze states in complex MR environments.

In sum, SPN emerges as a promising input for adaptive interfaces: not only as a signal of intent, but as a richer marker of anticipatory state that can guide feedback, disambiguate gaze, and support personalized interaction strategies.
In the following, we propose and discuss potential example use cases.

\subsubsection{Example Use Cases}

We follow prior HCI work that links sensing results to concrete interaction scenarios, such as VibroComm's vibroacoustic applications \cite{xiao2020vibrocomm} and Pose-on-the-Go's smartphone-based pose applications \cite{ahuja2021pose}. We outline three adaptive MR mechanisms enabled by our findings that SPN directly measures anticipatory uncertainty, varies continuously within individuals, and replicates across interface contexts.

\paragraph{Dynamic Dwell-Time Adjustment.}
Fixed dwell times force users to choose between speed and accuracy. Recent approaches adapt dwell based on gaze behavior: GazeIntent \cite{narkar2024gazeintent} predicts selection intent from fixation patterns and eye movement velocity, with personalized models preferred by 63\% of users. However, gaze behavior alone conflates multiple cognitive states, slow eye movements might indicate careful deliberation or simply tracking difficulty \cite{davidjohn2021towards}.

Our SPN measure resolves this ambiguity by revealing the user's internal certainty during the first 400ms of looking at a target. When the system detects high uncertainty (matching the pattern we observed during passive observation without feedback), it extends the required dwell time to prevent premature errors. When uncertainty is low (matching the pattern during confident selection with feedback), the system reduces dwell time to accelerate interaction. This approach distinguishes genuine hesitation from motor noise: a user looking slowly with high neural uncertainty needs more time to decide, while a user looking slowly with low uncertainty simply has slower eye movements. A practical implementation would combine gaze dynamics for rapid behavioral assessment with SPN for cognitive certainty, creating a two-layer adaptive system that responds within 400ms of fixation onset.

\paragraph{Uncertainty-Contingent Confirmation.}
Multimodal confirmation strategies, such as Gaze+Pinch \cite{pfeuffer2017gaze} and DualGaze \cite{mohan2018dualgaze}, are designed to reduce false positives but apply the same confirmation requirements regardless of user certainty. Systems either always require confirmation (eliminating errors but slowing interaction) or never require it (speeding interaction but allowing mistakes).

Our findings enable graduated confirmation that adapts to real-time cognitive state. When the system detects high confidence, selections execute immediately with a brief undo window, optimizing speed. Under moderate uncertainty, the system requires subtle confirmation, e.g., a brief sustained gaze or icon highlight, balancing prevention and efficiency. When uncertainty is high, explicit confirmation is required via controller press or extended dwell. This three-tier approach addresses a limitation Wolf et al. \cite{wolf2021gaze} identified in AR manual tasks: their system detected when users looked at one object while reaching for another, but could not assess confidence when gaze and action aligned correctly. Combining their gaze-hand coordination framework with our uncertainty measure enables warnings like "You appear uncertain—confirm this selection?" before executing potentially costly actions in sequential workflows.

\paragraph{Personalized Interaction Profiles.}
Current approaches in personalization adapt single dimensions, such as dwell time \cite{narkar2024gazeintent}, spatial ability \cite{zhao2025spatial}, or gender preferences \cite{severitt2025gaze}, using behavioral proxies. However, our classification results revealed substantial individual differences: person-dependent models achieved up to 97\% accuracy while person independent models reached only 69\%, indicating that neural uncertainty patterns vary systematically across individuals.

We propose characterizing users along three dimensions during brief calibration. First, baseline uncertainty disposition distinguishes users who naturally exhibit high monitoring activity during observation from those who show low anticipatory activity. Second, feedback sensitivity quantifies how much users rely on system confirmations to resolve uncertainty. Our results showed that some users reduced uncertainty with feedback, while others exhibited minimal change. Third, context specificity reveals whether uncertainty patterns generalize across interface types or vary by domain. These profiles enable tailored interfaces: users with low baseline uncertainty receive streamlined designs with minimal feedback, while users with high uncertainty receive guided interfaces with prominent confirmations and safety mechanisms. Users exhibiting context-specific patterns receive adaptive mode switching, which adjusts behavior across both familiar and novel interface elements. This person-centered approach, grounded in neural signatures, contrasts with group-centered rules based on demographics. Future implementations could track how profiles evolve with expertise and combine SPN with pupil dilation and electrodermal activity to monitor cognitive state multimodally across multiple timescales.

\subsection{Limitations and Future Work}

While our study demonstrates robust SPN effects in MR and their modulation by intention and feedback, its scope is not the validation of a complete MR interaction technique or user-facing UI adaptations. Instead, we focus on establishing SPN as an implicit neural marker of anticipatory states during ecologically valid gaze-based MR interaction. Accordingly, the limitations discussed below primarily concern how this signal can be further interpreted, disambiguated, and integrated into future MR systems.

First, our design did not directly probe how users respond when system feedback contradicts their expectations. A useful extension would be to include a control condition in which participants occasionally receive feedback unrelated to their actions. Presenting such unexpected feedback would allow testing for error-related potentials (ERN/Ne) in addition to SPN \cite{chavarriaga2010learning}. Beyond scientific interest, detecting error-related responses could provide practical value for MR interfaces by identifying when users recognize system misactivations, thus enabling automatic error correction or confidence calibration in gaze-based selection systems.

Second, while our results suggest that SPN reflects anticipatory uncertainty rather than pure intention, this ambiguity poses challenges for application. Distinguishing between "I am about to act" and "I am uncertain what will happen" is central for adaptive MR, yet both states can produce similar SPN responses. Future work should systematically vary both \textit{Intention} (Select vs. Observe) and \textit{Uncertainty} through controlled manipulations of system reliability or temporal predictability. Such studies would clarify whether SPN primarily reflects outcome unpredictability and whether clear intentions help stabilize anticipatory responses. Moreover, we acknowledge that, from a user’s subjective perspective, both conditions may appear similar, as sustained gaze leads to a system-side event in both cases, and the distinction between a UI action and a non-UI action may not always be consciously articulated.

Third, our study examined anticipation at the level of single fixations, whereas real MR interaction unfolds across sequences of revisits, comparisons, and scanpath motifs. SPN provides a robust index of pre-event anticipation, but its slow dynamics limit insights into how expectancy evolves across multi-step interaction flows \cite{eraslan2015eye, wollstadt2021quantifying}. Future work could integrate SPN analysis with scanpath-based measures from eye tracking to capture how anticipatory states develop across extended MR tasks. Such multimodal approaches would help bridge neural markers of anticipation with the temporal variance of natural MR gaze behavior.

Fourth, Select and Observe conditions shared identical sensorimotor demands (750ms gaze fixation). While this matching was intentional to isolate cognitive intention from motor confounds (by avoiding RP/LRP from button presses), it raises questions about whether participants genuinely maintained distinct internal goals. Three lines of evidence support the manipulation's validity: a statistically robust main effect of Intention, a significant Intention × Feedback interaction demonstrating differential processing of anticipated outcomes (if both were treated as selection, feedback should have modulated them equivalently), and high classification accuracy (75-97\%) achieved before any visual response. Nevertheless, future work could strengthen this through post-trial subjective ratings of intention clarity, think-aloud protocols, or complementary motor actions, though the latter would introduce motor preparation potentials that complicate SPN interpretation.

Fifth, although we report both person-dependent and person-independent decoding, these two settings were treated as separate modeling problems. Here, a next step is to explore hybrid approaches that combine population-level pretraining with efficient user-specific adaptation. Prior work in EEG transfer learning has shown that multi-user pretraining followed by user-level adaptation can substantially improve generalization across individuals \cite{he2020deep}. Rather than training a new model for each participant or relying solely on cross-subject learning, a unified multi-user model could first be trained to capture generalizable anticipatory EEG patterns and then adapted to a new user with minimal calibration data. Treating each user as a distinct domain aligns with recent advances in EEG domain adaptation, which highlight the need for feature alignment and subject-invariant representations \cite{lotte2018review, li2020review, zhao2019cross, xu2020subject}. Techniques such as lightweight fine-tuning, adapter modules, or feature-alignment methods could therefore provide a principled path toward rapid personalization. Such a strategy has the potential to increase robustness, reduce calibration time, and support more reliable deployment of intention-aware MR systems in real-world settings.

Finally, while this work focuses on neural and behavioral markers of anticipation, complementary qualitative methods such as interviews or post-task questionnaires could enrich future studies by capturing how users consciously interpret feedback, control, and system behavior. However, such methods are inherently limited in their ability to access pre-reflective anticipatory processes that unfold before overt action. We therefore view qualitative inquiry as a complementary layer that can contextualize neural markers, rather than as a substitute for measuring implicit cognitive states that users may not be able to reliably report.

\section{Conclusion}

We investigated how to decouple attention and intention in MR by leveraging gaze and EEG, addressing the core of the Midas Touch problem. Our results show that SPN is robustly elicited across everyday MR tasks, strongest under uncertainty, and shaped by the interaction of intention and feedback. This reframes SPN from a preparatory potential to a marker of anticipatory uncertainty with direct implications for adaptive interface design.  We also demonstrated that user intention can be reliably decoded with deep learning, achieving accuracies up to 97\% in person-dependent models. Together, these findings establish SPN as both a theoretical marker of anticipatory states and a practical signal for building intention-aware MR interfaces that reduce false activations and personalize interaction.

\section*{Open Science \& Transparency}
We encourage readers to reproduce and extend our results and analysis methods. Therefore, our experimental setup collected datasets, and analysis scripts are openly available on the Open Science Framework (\url{http://osf.io/9x6jy/}). During the preparation of this work, the authors used OpenAI’s GPT-5 and Grammarly for grammar and style editing. All content was reviewed and edited by the authors, who take full responsibility for the final publication.

\begin{acks}
This research was funded by the Deutsche Forschungsgemeinschaft (DFG, German Research Foundation) – Project-ID 251654672 – TRR 161 and under project number 529719707, "Multimodale Kommunikation von Intention in autonomen Systemen".
This work has been partly supported by the Research Center Trustworthy Data Science and Security (\href{https://rc-trust.ai}{https://rc-trust.ai}), one of the Research Alliance centers within the UA Ruhr (\href{https://uaruhr.de}{https://uaruhr.de}).  
\end{acks}

\bibliographystyle{ACM-Reference-Format}
\bibliography{bibliography}

@inproceedings{reddy2024towards,
author = {Reddy, G S Rajshekar and Proulx, Michael J and Hirshfield, Leanne and Ries, Anthony},
title = {Towards an Eye-Brain-Computer Interface: Combining Gaze with the Stimulus-Preceding Negativity for Target Selections in XR},
year = {2024},
isbn = {9798400703300},
publisher = {Association for Computing Machinery},
address = {New York, NY, USA},
url = {https://doi.org/10.1145/3613904.3641925},
doi = {10.1145/3613904.3641925},
booktitle = {Proceedings of the 2024 CHI Conference on Human Factors in Computing Systems},
articleno = {376},
numpages = {17},
location = {Honolulu, HI, USA},
series = {CHI '24}
}

@article{vanboxtel1994motor,
  title   = {Motor and non-motor aspects of slow brain potentials},
  author  = {van Boxtel, G. J. M. and Brunia, C. H. M.},
  journal = {Biological Psychology},
  volume  = {38},
  number  = {1},
  pages   = {37--51},
  year    = {1994},
  doi     = {10.1016/0301-0511(94)90048-5},
  publisher = {Elsevier}
}

@inproceedings{long2024detecting,
author = {Long, Xingyu and Mayer, Sven and Chiossi, Francesco},
title = {Multimodal Detection of External and Internal Attention in Virtual Reality using EEG and Eye Tracking Features},
year = {2024},
isbn = {9798400709982},
publisher = {Association for Computing Machinery},
address = {New York, NY, USA},
url = {https://doi.org/10.1145/3670653.3670657},
doi = {10.1145/3670653.3670657},
booktitle = {Proceedings of Mensch Und Computer 2024},
pages = {29–43},
numpages = {15},
location = {Karlsruhe, Germany},
series = {MuC '24}
}

@INPROCEEDINGS{chiossi2024mind,
  author={Chiossi, Francesco and Weiss, Yannick and Steinbrecher, Thomas and Mai, Christian and Kosch, Thomas},
  booktitle={2024 IEEE International Symposium on Mixed and Augmented Reality (ISMAR)}, 
  title={Mind the Visual Discomfort: Assessing Event-Related Potentials as Indicators for Visual Strain in Head-Mounted Displays}, 
  year={2024},
  volume={},
  number={},
  pages={150-159},
  doi={10.1109/ISMAR62088.2024.00029}}

@article{kohrs2016delays,
  title={Delays in human-computer interaction and their effects on brain activity},
  author={Kohrs, Christin and Angenstein, Nicole and Brechmann, Andre},
  journal={PLoS ONE},
  volume={11},
  number={1},
  pages={e0146250},
  year={2016},
  publisher={Public Library of Science},
  doi={10.1371/journal.pone.0146250},
  url={https://doi.org/10.1371/journal.pone.0146250}
}

@inproceedings{liu2014effects,
  title={The effects of interactive latency on exploratory visual analysis},
  author={Liu, Zhicheng and Heer, Jeffrey},
  booktitle={Proceedings of the 32nd annual ACM conference on Human factors in computing systems},
  pages={2349--2358},
  year={2014},
  organization={ACM},
  doi={10.1145/2556288.2557214},
  url={https://dl.acm.org/doi/10.1145/2556288.2557214}
}

@article{wang2009construction,
	title        = {The construction of a Williams design and randomization in cross-over clinical trials using SAS},
	author       = {Wang, Bing-Shun and Wang, Xiao-Jin and Gong, Li-Kun},
	year         = 2009,
	journal      = {Journal of statistical software},
	volume       = 29,
	pages        = {1--10},
	doi          = {10.18637/jss.v029.c01}
}

@article{tanovic2019anticipating,
  title     = {Anticipating the unknown: The stimulus-preceding negativity is enhanced by uncertain threat},
  author    = {Tanovic, Ema and Joormann, Jutta},
  journal   = {International Journal of Psychophysiology},
  volume    = {139},
  pages     = {68--73},
  year      = {2019},
  month     = {may},
  doi       = {10.1016/j.ijpsycho.2019.03.009},
  publisher = {Elsevier}
}

@article{shishkin2016eeg,
  title={EEG negativity in fixations used for gaze-based control: Toward converting intentions into actions with an eye-brain-computer interface},
  author={Shishkin, Sergei L and Nuzhdin, Yuri O and Svirin, Evgeny P and Trofimov, Alexander G and Fedorova, Anastasia A and Kozyrskiy, Bogdan L and Velichkovsky, Boris M},
  journal={Frontiers in neuroscience},
  volume={10},
  pages={528},
  year={2016},
  publisher={Frontiers Media SA},
 doi={https://doi.org/10.3389/fnins.2016.00528}
}

@article{lotte2018review,
  title        = {A review of classification algorithms for {EEG}-based brain--computer interfaces: a 10 year update},
  author       = {Lotte, Fabien and Bougrain, Laurent and Cichocki, Andrzej and Clerc, Maureen and Congedo, Marco and Rakotomamonjy, Alain and Yger, Florian},
  journal      = {Journal of Neural Engineering},
  volume       = {15},
  number       = {3},
  pages        = {031005},
  year         = {2018},
  publisher    = {IOP Publishing},
  doi          = {10.1088/1741-2552/aab2f2}
}

@article{schirrmeister2017deepeeg,
  title        = {Deep learning with convolutional neural networks for {EEG} decoding and visualization},
  author       = {Schirrmeister, Robin Tibor and Springenberg, Jost Tobias and Fiederer, Lukas Dominique Josef and Glasstetter, Martin and Eggensperger, Katharina and Tangermann, Michael and Hutter, Frank and Burgard, Wolfram and Ball, Tonio},
  journal      = {Human Brain Mapping},
  volume       = {38},
  number       = {11},
  pages        = {5391--5420},
  year         = {2017},
  publisher    = {Wiley},
  doi          = {10.1002/hbm.23730}
}

@article{chiossi2024understanding,
author = {Chiossi, Francesco and Gruenefeld, Uwe and Hou, Baosheng James and Newn, Joshua and Ou, Changkun and Liao, Rulu and Welsch, Robin and Mayer, Sven},
title = {Understanding the Impact of the Reality-Virtuality Continuum on Visual Search Using Fixation-Related Potentials and Eye Tracking Features},
year = {2024},
issue_date = {September 2024},
publisher = {Association for Computing Machinery},
address = {New York, NY, USA},
volume = {8},
number = {MHCI},
url = {https://doi.org/10.1145/3676528},
doi = {10.1145/3676528},
journal = {Proc. ACM Hum.-Comput. Interact.},
month = sep,
articleno = {281},
numpages = {33}
}

@ARTICLE{chiossi2024searching,
  author={Chiossi, Francesco and Trautmannsheimer, Ines and Ou, Changkun and Gruenefeld, Uwe and Mayer, Sven},
  journal={IEEE Transactions on Visualization and Computer Graphics}, 
  title={Searching Across Realities: Investigating ERPs and Eye-Tracking Correlates of Visual Search in Mixed Reality}, 
  year={2024},
  volume={30},
  number={11},
  pages={6997-7007},
  doi={10.1109/TVCG.2024.3456172}
}

@article{faul2009statistical,
  title={Statistical power analyses using G* Power 3.1: Tests for correlation and regression analyses},
  author={Faul, Franz and Erdfelder, Edgar and Buchner, Axel and Lang, Albert-Georg},
  journal={Behavior research methods},
  volume={41},
  number={4},
  pages={1149--1160},
  year={2009},
  doi={10.3758/BRM.41.4.1149},
  publisher={Springer}
}

@incollection{yatani2016effect,
  author    = {Koji Yatani},
  title     = {Effect Sizes and Power Analysis in {HCI}},
  booktitle = {Modern Statistical Methods for {HCI}},
  editor    = {Judy Robertson and Maurits Kaptein},
  pages     = {87--110},
  year      = {2016},
  publisher = {Springer},
  address   = {Cham},
  doi       = {10.1007/978-3-319-26633-6\_5},
  url       = {https://doi.org/10.1007/978-3-319-26633-6_5}
}

@article{poli2007spn,
  title        = {Stimulus-Preceding Negativity and heart rate changes in anticipation of affective pictures},
  author       = {Poli, Silvia and Sarlo, Michela and Bortoletto, Marta and Buodo, Giulia and Palomba, Daniela},
  journal      = {International Journal of Psychophysiology},
  volume       = {65},
  number       = {1},
  pages        = {32--39},
  year         = {2007},
  publisher    = {Elsevier},
  doi          = {10.1016/j.ijpsycho.2007.02.008}
}

@inproceedings{rasch2025ar,
author = {Rasch, Julian and Wilhalm, Matthias and M\"{u}ller, Florian and Chiossi, Francesco},
title = {AR You on Track? Investigating Effects of Augmented Reality Anchoring on Dual-Task Performance While Walking},
year = {2025},
isbn = {9798400713941},
publisher = {Association for Computing Machinery},
address = {New York, NY, USA},
url = {https://doi.org/10.1145/3706598.3714258},
doi = {10.1145/3706598.3714258},
booktitle = {Proceedings of the 2025 CHI Conference on Human Factors in Computing Systems},
articleno = {1217},
numpages = {21},
location = {
},
series = {CHI '25}
}

@article{lawhern2018eegnet,
  title={EEGNet: a compact convolutional neural network for EEG-based brain--computer interfaces},
  author={Lawhern, Vernon J and Solon, Amelia J and Waytowich, Nicholas R and Gordon, Stephen M and Hung, Chou P and Lance, Brent J},
  journal={Journal of neural engineering},
  volume={15},
  number={5},
  pages={056013},
  year={2018},
  publisher={iOP Publishing}
}

@ARTICLE{santamaria2020inception,
  author={Santamaría-Vázquez, Eduardo and Martínez-Cagigal, Víctor and Vaquerizo-Villar, Fernando and Hornero, Roberto},
  journal={IEEE Transactions on Neural Systems and Rehabilitation Engineering}, 
  title={EEG-Inception: A Novel Deep Convolutional Neural Network for Assistive ERP-Based Brain-Computer Interfaces}, 
  year={2020},
  volume={28},
  number={12},
  pages={2773-2782},
  doi={10.1109/TNSRE.2020.3048106}
}

@inproceedings{akiba2019optuna,
  title={Optuna: A next-generation hyperparameter optimization framework},
  author={Akiba, Takuya and Sano, Shotaro and Yanase, Toshihiko and Ohta, Takeru and Koyama, Masanori},
  booktitle={Proceedings of the 25th ACM SIGKDD international conference on knowledge discovery \& data mining},
  pages={2623--2631},
  year={2019}
}

@article{chen2025steady,
  title={Steady-State Visual-Evoked-Potential--Driven Quadrotor Control Using a Deep Residual CNN for Short-Time Signal Classification},
  author={Chen, Jiannan and Yang, Chenju and Wei, Rao and Hua, Changchun and Mu, Dianrui and Sun, Fuchun},
  journal={Sensors},
  volume={25},
  number={15},
  pages={4779},
  year={2025},
  publisher={MDPI}
}

@article{schirrmeister2017deep,
  title={Deep learning with convolutional neural networks for EEG decoding and visualization},
  author={Schirrmeister, Robin Tibor and Springenberg, Jost Tobias and Fiederer, Lukas Dominique Josef and Glasstetter, Martin and Eggensperger, Katharina and Tangermann, Michael and Hutter, Frank and Burgard, Wolfram and Ball, Tonio},
  journal={Human brain mapping},
  volume={38},
  number={11},
  pages={5391--5420},
  year={2017},
  publisher={Wiley Online Library}
}

@article{craik2019deep,
  title={Deep learning for electroencephalogram (EEG) classification tasks: a review},
  author={Craik, Alexander and He, Yongtian and Contreras-Vidal, Jose L},
  journal={Journal of neural engineering},
  volume={16},
  number={3},
  pages={031001},
  year={2019},
  publisher={IOP Publishing}
}

@article{chavarriaga2010learning,
  title={Learning from EEG error-related potentials in noninvasive brain-computer interfaces},
  author={Chavarriaga, Ricardo and Mill{\'a}n, Jos{\'e} del R},
  journal={IEEE transactions on neural systems and rehabilitation engineering},
  volume={18},
  number={4},
  pages={381--388},
  year={2010},
  publisher={IEEE}
}

@article{moris2013learning,
  title={Learning-induced modulations of the stimulus-preceding negativity},
  author={Mor{\'\i}s, Joaqu{\'\i}n and Luque, David and Rodr{\'\i}guez-Fornells, Antoni},
  journal={Psychophysiology},
  volume={50},
  number={9},
  pages={931--939},
  year={2013},
  publisher={Wiley Online Library}
}

@article{jas2017autoreject,
  title={Autoreject: Automated artifact rejection for MEG and EEG data},
  author={Jas, Mainak and Engemann, Denis A. and Bekhti, Yousra and Raimondo, Federico and Gramfort, Alexandre},
  journal={NeuroImage},
  volume={159},
  pages={417--429},
  year={2017},
  publisher={Elsevier},
  doi={10.1016/j.neuroimage.2017.06.030},
  url={https://doi.org/10.1016/j.neuroimage.2017.06.030}
}

@article{poston_human_2000,
	title = {Human engineering design data digest},
	journal = {Washington, DC: Department of Defense Human Factors Engineering Technical Advisory Group},
	author = {Poston, Alan},
	year = {2000},
	pages = {61--75},
}

@article{hussain2023effects,
  title        = {Effects of Interaction Method, Size, and Distance to Object on Augmented Reality Interfaces},
  author       = {Hussain, Muhammad and Park, Jaehyun and Kim, Hyun K},
  journal      = {Interacting with Computers},
  volume       = {35},
  number       = {1},
  pages        = {1--11},
  year         = {2023},
  month        = {jan},
  doi          = {10.1093/iwc/iwad034},
  publisher    = {Oxford University Press},
  url          = {https://doi.org/10.1093/iwc/iwad034}
}

@article{thaler2013fixation,
  title        = {What is the best fixation target? The effect of target shape on stability of fixational eye movements},
  author       = {Thaler, Lore and Sch{\"u}tz, Alexander C. and Goodale, Melvyn A. and Gegenfurtner, Karl R.},
  journal      = {Vision Research},
  volume       = {76},
  pages        = {31--42},
  year         = {2013},
  publisher    = {Elsevier},
  doi          = {10.1016/j.visres.2012.10.012}
}

@article{li2022mne,
	title        = {MNE-ICALabel: Automatically annotating ICA components with ICLabel in Python},
	author       = {Li, Adam and Feitelberg, Jacob and Saini, Anand Prakash and H{\"o}chenberger, Richard and Scheltienne, Mathieu},
	year         = 2022,
	journal      = {Journal of Open Source Software},
	volume       = 7,
	number       = 76,
	pages        = 4484,
	doi          = {10.21105/joss.04484}
}

@article{lee1999independent,
	title        = {Independent component analysis using an extended infomax algorithm for mixed subgaussian and supergaussian sources},
	author       = {Lee, Te-Won and Girolami, Mark and Sejnowski, Terrence J},
	year         = 1999,
	journal      = {Neural computation},
	publisher    = {MIT Press},
address={Cambridge, MA, USA},
	volume       = 11,
	number       = 2,
	pages        = {417--441},
        doi={10.1162/089976699300016719}
}

@article{perrin1989spherical,
	title        = {Spherical splines for scalp potential and current density mapping},
	author       = {Perrin, Fran{\c{c}}ois and Pernier, Jacques and Bertrand, Olivier and Echallier, Jean Francois},
	year         = 1989,
	journal      = {Electroencephalography and clinical neurophysiology},
	publisher    = {Elsevier},
	volume       = 72,
	number       = 2,
	pages        = {184--187},
	doi          = {10.1016/0013-4694(89)90180-6}
}

@article{bigdely2015prep,
	title        = {The PREP pipeline: standardized preprocessing for large-scale EEG analysis},
	author       = {Bigdely-Shamlo, Nima and Mullen, Tim and Kothe, Christian and Su, Kyung-Min and Robbins, Kay A},
	year         = 2015,
	journal      = {Frontiers in neuroinformatics},
	publisher    = {Frontiers Media SA},
	volume       = 9,
	pages        = 16,
	doi          = {10.3389/fninf.2015.00016}
}

@article{gramfort2013meg,
	title        = {MEG and EEG data analysis with MNE-Python},
	author       = {Gramfort, Alexandre and Luessi, Martin and Larson, Eric and Engemann, Denis A and Strohmeier, Daniel and Brodbeck, Christian and Goj, Roman and Jas, Mainak and Brooks, Teon and Parkkonen, Lauri},
	year         = 2013,
	journal      = {Frontiers in neuroscience},
	publisher    = {Frontiers},
	pages        = 267,
        volume={7},
	doi          = {10.3389/fnins.2013.00267}
}

@article{atkinson1989and,
	title        = {‘Where’and ‘what’in visual search},
	author       = {Atkinson, Janette and Braddick, Oliver J},
	year         = 1989,
	journal      = {Perception},
	publisher    = {SAGE Publications Sage UK: London, England},
	volume       = 18,
	number       = 2,
	pages        = {181--189},
	doi          = {10.1068/p180181}
}

@article{woodman1999electrophysiological,
	title        = {Electrophysiological measurement of rapid shifts of attention during visual search},
	author       = {Woodman, Geoffrey F and Luck, Steven J},
	year         = 1999,
	journal      = {Nature},
	publisher    = {Nature Publishing Group UK London},
	volume       = 400,
	number       = 6747,
	pages        = {867--869},
	doi          = {10.1038/23698}
}

@article{liu2021app,
  title={How do app icon color and border shape influence visual search efficiency and user experience? Evidence from an eye-tracking study},
  author={Liu, Weilin and Cao, Yaqin and Proctor, Robert W},
  journal={International Journal of Industrial Ergonomics},
  volume={84},
  pages={103160},
  year={2021},
  publisher={Elsevier}, 
  doi = {10.1016/j.ergon.2021.103160}
}

@article{boudewyn2018many,
  title={How many trials does it take to get a significant ERP effect? It depends},
  author={Boudewyn, Megan A and Luck, Steven J and Farrens, Jaclyn L and Kappenman, Emily S},
  journal={Psychophysiology},
  volume={55},
  number={6},
  pages={e13049},
  year={2018},
  publisher={Wiley},
  doi={10.1111/psyp.13049}
}

@article{jensen2022towards,
  title={Towards thoughtful planning of ERP studies: How participants, trials, and effect sizes influence statistical power},
  author={Jensen, Darin and MacDonald, Kevin},
  journal={Psychophysiology},
  volume={59},
  number={12},
  pages={e14208},
  year={2022},
  publisher={Wiley},
  doi={10.1111/psyp.14245}
}

@article{combrisson2015exceeding,
  title={Exceeding chance level by chance: The caveat of theoretical chance levels in brain signal classification and statistical assessment of decoding accuracy},
  author={Combrisson, Etienne and Jerbi, Karim},
  journal={Journal of neuroscience methods},
  volume={250},
  pages={126--136},
  year={2015},
  publisher={Elsevier}
}

@article{he2020deep,
  title={Deep transfer learning for EEG-based brain–computer interfaces},
  author={He, Haibo and Wu, Dongrui},
  journal={IEEE Transactions on Neural Systems and Rehabilitation Engineering},
  volume={28},
  number={10},
  pages={2063--2073},
  year={2020}
}

@inproceedings{zhao2019cross,
  title={Cross-subject EEG emotion recognition based on domain adaptation and graph regularization},
  author={Zhao, Guofei and others},
  booktitle={International Joint Conference on Artificial Intelligence (IJCAI)},
  year={2019}
}

@article{xu2020subject,
  title={Subject-invariant representation learning for EEG-based emotion recognition via adversarial domain generalization},
  author={Xu, Jun and others},
  journal={IEEE Transactions on Affective Computing},
  year={2020}
}

@article{li2020review,
  title={A review of domain adaptation methods for EEG-based emotion recognition},
  author={Li, Yuming and Zheng, Wei-Long and Wang, Yijun and Lu, Bao-Liang},
  journal={Frontiers in Neuroscience},
  volume={14},
  pages={109},
  year={2020}
}

@article{tibshirani1993introduction,
  title={An introduction to the bootstrap},
  author={Tibshirani, Robert J and Efron, Bradley},
  journal={Monographs on statistics and applied probability},
  volume={57},
  number={1},
  pages={1--436},
  year={1993}
}

@article{clayson2019methodological,
  title={Methodological reporting behavior, sample sizes, and statistical power in studies of event-related potentials: Barriers to reproducibility and replicability},
  author={Clayson, Peter E and Carbine, Kaylie A and Baldwin, Scott A and Larson, Michael J},
  journal={Psychophysiology},
  volume={56},
  number={11},
  pages={e13437},
  year={2019},
  publisher={Wiley Online Library}, 
   doi={ 10.1111/psyp.13437}
}

@article{damen1994spn,
  author    = {Damen, Ernst J. P. and Brunia, Cornelis H. M.},
  title     = {Is a stimulus conveying task-relevant information a sufficient condition to elicit a stimulus-preceding negativity?},
  journal   = {Psychophysiology},
  volume    = {31},
  number    = {2},
  pages     = {129--139},
  year      = {1994},
  month     = mar,
  doi       = {10.1111/j.1469-8986.1994.tb01033.x},
  publisher = {Wiley}
}

@incollection{brunia2012negativeslow,
  author    = {Brunia, Cornelis H. M. and van Boxtel, Geert J. M. and B{\"o}cker, Koen B. E.},
  title     = {Negative Slow Waves as Indices of Anticipation: The Bereitschaftspotential, the Contingent Negative Variation, and the Stimulus-Preceding Negativity},
  booktitle = {The Oxford Handbook of Event-Related Potential Components},
  editor    = {Kappenman, Emily S. and Luck, Steven J.},
  pages     = {190--208},
  year      = {2012},
  publisher = {Oxford University Press},
  address   = {Oxford, UK},
  doi       = {10.1093/oxfordhb/9780195374148.013.0108}
}

@article{shibasaki2006bereitschaftspotential,
  author    = {Shibasaki, Hiroshi and Hallett, Mark},
  title     = {What is the Bereitschaftspotential?},
  journal   = {Clinical Neurophysiology},
  volume    = {117},
  number    = {11},
  pages     = {2341--2356},
  year      = {2006},
  month     = nov,
  doi       = {10.1016/j.clinph.2006.04.025},
  publisher = {Elsevier}
}

@article{libet1983preparation,
  author    = {Libet, Benjamin and Wright, Elwood W. and Gleason, Curtis A.},
  title     = {Preparation- or intention-to-act, in relation to pre-event potentials recorded at the vertex},
  journal   = {Electroencephalography and Clinical Neurophysiology},
  volume    = {56},
  number    = {4},
  pages     = {367--372},
  year      = {1983},
  month     = oct,
  doi       = {10.1016/0013-4694(83)90281-x},
  publisher = {Elsevier}
}

@inproceedings{pfeuffer2017gaze,
author = {Pfeuffer, Ken and Mayer, Benedikt and Mardanbegi, Diako and Gellersen, Hans},
title = {Gaze + pinch interaction in virtual reality},
year = {2017},
isbn = {9781450354868},
publisher = {Association for Computing Machinery},
address = {New York, NY, USA},
url = {https://doi.org/10.1145/3131277.3132180},
doi = {10.1145/3131277.3132180},
booktitle = {Proceedings of the 5th Symposium on Spatial User Interaction},
pages = {99–108},
numpages = {10},
location = {Brighton, United Kingdom},
series = {SUI '17}
}

@inproceedings{lystbaek2024hands,
author = {Lystb\ae{}k, Mathias N. and Mikkelsen, Thorbj\o{}rn and Krisztandl, Roland and Gonzalez, Eric J and Gonzalez-Franco, Mar and Gellersen, Hans and Pfeuffer, Ken},
title = {Hands-on, Hands-off: Gaze-Assisted Bimanual 3D Interaction},
year = {2024},
isbn = {9798400706288},
publisher = {Association for Computing Machinery},
address = {New York, NY, USA},
url = {https://doi.org/10.1145/3654777.3676331},
doi = {10.1145/3654777.3676331},
booktitle = {Proceedings of the 37th Annual ACM Symposium on User Interface Software and Technology},
articleno = {80},
numpages = {12},
location = {Pittsburgh, PA, USA},
series = {UIST '24}
}

@inproceedings{jinwook2025pinch,
author = {Kim, Jinwook and Park, Sangmin and Zhou, Qiushi and Gonzalez-Franco, Mar and Lee, Jeongmi and Pfeuffer, Ken},
title = {PinchCatcher: Enabling Multi-selection for Gaze+Pinch},
year = {2025},
isbn = {9798400713941},
publisher = {Association for Computing Machinery},
address = {New York, NY, USA},
url = {https://doi.org/10.1145/3706598.3713530},
doi = {10.1145/3706598.3713530},
booktitle = {Proceedings of the 2025 CHI Conference on Human Factors in Computing Systems},
articleno = {853},
numpages = {16},
location = {
},
series = {CHI '25}
}

@inproceedings{jacob1990what,
author = {Jacob, Robert J. K.},
title = {What you look at is what you get: eye movement-based interaction techniques},
year = {1990},
isbn = {0201509326},
publisher = {Association for Computing Machinery},
address = {New York, NY, USA},
url = {https://doi.org/10.1145/97243.97246},
doi = {10.1145/97243.97246},
booktitle = {Proceedings of the SIGCHI Conference on Human Factors in Computing Systems},
pages = {11–18},
numpages = {8},
location = {Seattle, Washington, USA},
series = {CHI '90}
}

@article{burnham2025effects,
  title={The effects of dynamic dwell time systems on the usability of eye-tracking technology: a systematic review and meta-analyses},
  author={Burnham, Seamus PL and Bonar, Sonja and Collins, Mackenzie and Davies, T Claire},
  journal={Human--Computer Interaction},
  pages={1--25},
  year={2025},
  publisher={Taylor \& Francis}
}

@inproceedings{kyto2018pipointing,
author = {Kyt\"{o}, Mikko and Ens, Barrett and Piumsomboon, Thammathip and Lee, Gun A. and Billinghurst, Mark},
title = {Pinpointing: Precise Head- and Eye-Based Target Selection for Augmented Reality},
year = {2018},
isbn = {9781450356206},
publisher = {Association for Computing Machinery},
address = {New York, NY, USA},
url = {https://doi.org/10.1145/3173574.3173655},
doi = {10.1145/3173574.3173655},
booktitle = {Proceedings of the 2018 CHI Conference on Human Factors in Computing Systems},
pages = {1–14},
numpages = {14},
location = {Montreal QC, Canada},
series = {CHI '18}
}

@inproceedings{mutasim2021pinch,
author = {Mutasim, Aunnoy K and Batmaz, Anil Ufuk and Stuerzlinger, Wolfgang},
title = {Pinch, Click, or Dwell: Comparing Different Selection Techniques for Eye-Gaze-Based Pointing in Virtual Reality},
year = {2021},
isbn = {9781450383455},
publisher = {Association for Computing Machinery},
address = {New York, NY, USA},
url = {https://doi.org/10.1145/3448018.3457998},
doi = {10.1145/3448018.3457998},
booktitle = {ACM Symposium on Eye Tracking Research and Applications},
articleno = {15},
numpages = {7},
location = {Virtual Event, Germany},
series = {ETRA '21 Short Papers}
}

@inproceedings{lu2021itext,
author = {Lu, Xueshi and Yu, Difeng and Liang, Hai-Ning and Goncalves, Jorge},
title = {iText: Hands-free Text Entry on an Imaginary Keyboard for Augmented Reality Systems},
year = {2021},
isbn = {9781450386357},
publisher = {Association for Computing Machinery},
address = {New York, NY, USA},
url = {https://doi.org/10.1145/3472749.3474788},
doi = {10.1145/3472749.3474788},
booktitle = {The 34th Annual ACM Symposium on User Interface Software and Technology},
pages = {815–825},
numpages = {11},
location = {Virtual Event, USA},
series = {UIST '21}
}

@inproceedings{protzak2013passivebci,
  author    = {Protzak, Janna and Ihme, Klas and Zander, Thorsten Oliver},
  title     = {A Passive Brain-Computer Interface for Supporting Gaze-Based Human-Machine Interaction},
  booktitle = {Universal Access in Human-Computer Interaction. Design Methods, Tools, and Interaction Techniques for eInclusion: 7th International Conference, UAHCI 2013, Held as Part of HCI International 2013, Las Vegas, NV, USA, July 21-26, 2013, Proceedings, Part I},
  series    = {Lecture Notes in Computer Science},
  volume    = {8009},
  pages     = {662--671},
  year      = {2013},
  publisher = {Springer},
  doi       = {10.1007/978-3-642-39188-0_71}
}

@article{vrieze2012aicbic,
  author    = {Vrieze, Scott I.},
  title     = {Model selection and psychological theory: A discussion of the differences between the Akaike information criterion (AIC) and the Bayesian information criterion (BIC)},
  journal   = {Psychological Methods},
  volume    = {17},
  number    = {2},
  pages     = {228--243},
  year      = {2012},
  doi       = {10.1037/a0027127},
  publisher = {American Psychological Association}
}

@article{ries2018Lambda,
  title        = {The fixation-related lambda response: Effects of saccade magnitude, spatial frequency, and ocular artifact removal},
  author       = {Ries, Anthony J. and Slayback, David and Touryan, Jon},
  journal      = {International Journal of Psychophysiology},
  volume       = {134},
  pages        = {1--8},
  year         = {2018},
  doi          = {10.1016/j.ijpsycho.2018.09.004},
  url          = {https://doi.org/10.1016/j.ijpsycho.2018.09.004},
  issn         = {0167-8760},
  publisher    = {Elsevier},
}

@article{syrov2024visuomotor,
  author    = {Syrov, Nikolay and Yakovlev, Lev and Kaplan, Alexander and Lebedev, Mikhail},
  title     = {Motor cortex activation during visuomotor transformations: evoked potentials during overt and imagined movements},
  journal   = {Cerebral Cortex},
  volume    = {34},
  number    = {1},
  pages     = {bhad440},
  year      = {2024},
  month     = jan,
  doi       = {10.1093/cercor/bhad440},
  publisher = {Oxford University Press}
}

@article{kotani2003spn,
  author    = {Kotani, Yasunori and Kishida, Sachiko and Hiraku, Shiho and Suda, Kazuhiro and Ishii, Motonobu and Aihara, Yasutsugu},
  title     = {Effects of information and reward on stimulus-preceding negativity prior to feedback stimuli},
  journal   = {Psychophysiology},
  volume    = {40},
  number    = {5},
  pages     = {818--826},
  year      = {2003},
  month     = sep,
  doi       = {10.1111/1469-8986.00082},
  publisher = {Wiley}
}

@article{pfeuffer2024gazepinch,
  author    = {Pfeuffer, Ken and Gellersen, Hans and Gonzalez-Franco, Mar},
  title     = {Design Principles and Challenges for Gaze + Pinch Interaction in XR},
  journal   = {IEEE Computer Graphics and Applications},
  volume    = {44},
  number    = {3},
  pages     = {74--81},
  year      = {2024},
  month     = may,
  doi       = {10.1109/MCG.2024.3382961},
  publisher = {IEEE}
}

@inproceedings{wagner2023fitts,
author = {Wagner, Uta and Lystb\ae{}k, Mathias N. and Manakhov, Pavel and Gr\o{}nb\ae{}k, Jens Emil Sloth and Pfeuffer, Ken and Gellersen, Hans},
title = {A Fitts’ Law Study of Gaze-Hand Alignment for Selection in 3D User Interfaces},
year = {2023},
isbn = {9781450394215},
publisher = {Association for Computing Machinery},
address = {New York, NY, USA},
url = {https://doi.org/10.1145/3544548.3581423},
doi = {10.1145/3544548.3581423},
booktitle = {Proceedings of the 2023 CHI Conference on Human Factors in Computing Systems},
articleno = {252},
numpages = {15},
location = {Hamburg, Germany},
series = {CHI '23}
}

@article{horstmann2005target,
  author    = {Horstmann, Annette and Hoffmann, Klaus-Peter},
  title     = {Target selection in eye--hand coordination: Do we reach to where we look or do we look to where we reach?},
  journal   = {Experimental Brain Research},
  volume    = {167},
  pages     = {187--195},
  year      = {2005},
  month     = jul,
  doi       = {10.1007/s00221-005-0038-z},
  publisher = {Springer}
}

@inproceedings{hirzle2020survey,
author = {Hirzle, Teresa and Cordts, Maurice and Rukzio, Enrico and Bulling, Andreas},
title = {A Survey of Digital Eye Strain in Gaze-Based Interactive Systems},
year = {2020},
isbn = {9781450371339},
publisher = {Association for Computing Machinery},
address = {New York, NY, USA},
url = {https://doi.org/10.1145/3379155.3391313},
doi = {10.1145/3379155.3391313},
booktitle = {ACM Symposium on Eye Tracking Research and Applications},
articleno = {9},
numpages = {12},
location = {Stuttgart, Germany},
series = {ETRA '20 Full Papers}
}

@inproceedings{pfeuffer2020empirical,
author = {Pfeuffer, Ken and Mecke, Lukas and Delgado Rodriguez, Sarah and Hassib, Mariam and Maier, Hannah and Alt, Florian},
title = {Empirical Evaluation of Gaze-enhanced Menus in Virtual Reality},
year = {2020},
isbn = {9781450376198},
publisher = {Association for Computing Machinery},
address = {New York, NY, USA},
url = {https://doi.org/10.1145/3385956.3418962},
doi = {10.1145/3385956.3418962},
booktitle = {Proceedings of the 26th ACM Symposium on Virtual Reality Software and Technology},
articleno = {20},
numpages = {11},
location = {Virtual Event, Canada},
series = {VRST '20}
}

@article{zavichi2025gazehand,
  author    = {Zavichi, Mona and Santos, Andr{\'e} and Moreira, Catarina and Maciel, Anderson and Jorge, Joaquim},
  title     = {Gaze--Hand Steering for Travel and Multitasking in Virtual Environments},
  journal   = {Multimodal Technologies and Interaction},
  volume    = {9},
  number    = {6},
  pages     = {61},
  year      = {2025},
  month     = jun,
  doi       = {10.3390/mti9060061},
  publisher = {MDPI}
}

@inproceedings{kirst2016verge,
  author    = {Kirst, Dominik and Bulling, Andreas},
  title     = {On the Verge: Voluntary Convergences for Accurate and Precise Timing of Gaze Input},
  booktitle = {Proceedings of the 2016 CHI Conference Extended Abstracts on Human Factors in Computing Systems (CHI EA '16)},
  pages     = {1519--1525},
  address   = {San Jose, California, USA},
  year      = {2016},
  publisher = {Association for Computing Machinery},
  doi       = {10.1145/2851581.2892307}
}

@article{gehrke2025,
  title={Modeling the Intent to Interact With VR Using Physiological Features},
  author={Nguyen, Willy and Gramann, Klaus and Gehrke, Lukas},
  journal={IEEE Transactions on Visualization and Computer Graphics},
  volume={30},
  number={8},
  pages={5893--5902},
  year={2024},
  doi={10.1109/TVCG.2023.3308787}
}

@inproceedings{park2014,
  title={Human Implicit Intent Discrimination Using EEG and Eye Movement},
  author={Park, Ukeob and Mallipeddi, Rammohan and Lee, Minho},
  booktitle={International Conference on Neural Information Processing (ICONIP)},
  series={Lecture Notes in Computer Science},
  volume={8834},
  pages={11--18},
  year={2014},
  publisher={Springer},
  doi={10.1007/978-3-319-12637-1_2}
}

@article{kang2015implicit,
  title={Human implicit intent recognition based on the phase synchrony of EEG signals},
  author={Kang, Jun-Su and Park, Ukeob and Gonuguntla, V. and Veluvolu, K.C. and Lee, Minho},
  journal={Pattern Recognition Letters},
  volume={66},
  pages={144--152},
  year={2015},
  publisher={Elsevier},
  doi={10.1016/j.patrec.2015.06.013}
}

@ARTICLE{zhang2020making,
  author={Zhang, Dalin and Yao, Lina and Chen, Kaixuan and Wang, Sen and Chang, Xiaojun and Liu, Yunhao},
  journal={IEEE Transactions on Cybernetics}, 
  title={Making Sense of Spatio-Temporal Preserving Representations for EEG-Based Human Intention Recognition}, 
  year={2020},
  volume={50},
  number={7},
  pages={3033-3044},
  doi={10.1109/TCYB.2019.2905157}}

@article{jackson1997behavioral,
  author    = {Jackson, Cynthia M. and Chow, Simeon and Leitch, Robert A.},
  title     = {Toward an Understanding of the Behavioral Intention to Use an Information System},
  journal   = {Decision Sciences},
  volume    = {28},
  number    = {2},
  pages     = {357--389},
  year      = {1997},
  month     = jun,
  doi       = {10.1111/j.1540-5915.1997.tb01315.x},
  publisher = {Wiley}
}

@article{villa2025agency,
  author    = {Villa, Steeven and Barth, Lisa L. and Chiossi, Francesco and Welsch, Robin and Kosch, Thomas},
  title     = {Whose mind is it anyway? A systematic review and exploration on agency in cognitive augmentation},
  journal   = {Computers in Human Behavior: Artificial Humans},
  volume    = {5},
  pages     = {100158},
  year      = {2025},
  month     = aug,
  doi       = {10.1016/j.chbah.2025.100158},
  publisher = {Elsevier}
}

@article{delnatte2023freewill,
  author    = {Delnatte, Claire and Roze, Emmanuel and Pouget, Pierre and Gall{\'e}a, C{\'e}cile and Welniarz, Quentin},
  title     = {Can neuroscience enlighten the philosophical debate about free will?},
  journal   = {Neuropsychologia},
  volume    = {185},
  pages     = {108632},
  year      = {2023},
  doi       = {10.1016/j.neuropsychologia.2023.108632},
  publisher = {Elsevier}
}

@inproceedings{sharma2024distinguishing,
author = {Sharma, Mansi and Mart\'{\i}nez Mart\'{\i}nez, Camilo Andr\'{e}s and Wirth, Benedikt Emanuel and Kr\"{u}ger, Antonio and M\"{u}ller, Philipp},
title = {Distinguishing Target and Non-Target Fixations with EEG and Eye Tracking in Realistic Visual Scenes},
year = {2024},
isbn = {9798400704628},
publisher = {Association for Computing Machinery},
address = {New York, NY, USA},
url = {https://doi.org/10.1145/3678957.3685728},
doi = {10.1145/3678957.3685728},
booktitle = {Proceedings of the 26th International Conference on Multimodal Interaction},
pages = {459–468},
numpages = {10},
location = {San Jose, Costa Rica},
series = {ICMI '24}
}

@inproceedings{park2024impact,
author = {Park, Yeji and Kim, Jiwan and Oakley, Ian},
title = {The Impact of Gaze and Hand Gesture Complexity on Gaze-Pinch Interaction Performances},
year = {2024},
isbn = {9798400710582},
publisher = {Association for Computing Machinery},
address = {New York, NY, USA},
url = {https://doi.org/10.1145/3675094.3678990},
doi = {10.1145/3675094.3678990},
booktitle = {Companion of the 2024 on ACM International Joint Conference on Pervasive and Ubiquitous Computing},
pages = {622–626},
numpages = {5},
location = {Melbourne VIC, Australia},
series = {UbiComp '24}
}

@article{mrotek2007target,
  author    = {Mrotek, Leigh A. and Soechting, John F.},
  title     = {Target Interception: Hand--Eye Coordination and Strategies},
  journal   = {Journal of Neuroscience},
  volume    = {27},
  number    = {27},
  pages     = {7297--7309},
  year      = {2007},
  month     = jul,
  doi       = {10.1523/JNEUROSCI.2046-07.2007},
  publisher = {Society for Neuroscience}
}

@inproceedings{klamka2015lookpedal,
  author    = {Klamka, Konstantin and Siegel, Andreas and Vogt, Stefan and G{\"o}bel, Fabian and Stellmach, Sophie and Dachselt, Raimund},
  title     = {Look \& Pedal: Hands-Free Navigation in Zoomable Information Spaces through Gaze-Supported Foot Input},
  booktitle = {Proceedings of the 2015 ACM International Conference on Multimodal Interaction (ICMI '15)},
  year      = {2015},
  pages     = {123--130},
  address   = {Seattle, Washington, USA},
  publisher = {Association for Computing Machinery},
  doi       = {10.1145/2818346.2820751}
}

@article{lee2014ar,
  title     = {Design and implementation of an augmented reality system using gaze interaction},
  author    = {Lee, Jae-Young and Park, Hyung-Min and Lee, Seok-Han and Shin, Soon-Ho and Kim, Tae-Eun and Choi, Jong-Soo},
  journal   = {Multimedia Tools and Applications},
  volume    = {68},
  pages     = {265--280},
  year      = {2014},
  month     = dec,
  doi       = {10.1007/s11042-011-0864-2},
  publisher = {Springer}
}

@INPROCEEDINGS{whithlock2018interacting,
  author={Whitlock, Matt and Harnner, Ethan and Brubaker, Jed R. and Kane, Shaun and Szafir, Danielle Albers},
  booktitle={2018 IEEE Conference on Virtual Reality and 3D User Interfaces (VR)}, 
  title={Interacting with Distant Objects in Augmented Reality}, 
  year={2018},
  volume={},
  number={},
  pages={41-48},
  doi={10.1109/VR.2018.8446381}
}

@article{deecke1969readiness,
  author    = {Deecke, L. and Scheid, P. and Kornhuber, H. H.},
  title     = {Distribution of readiness potential, pre-motion positivity, and motor potential of the human cerebral cortex preceding voluntary finger movements},
  journal   = {Experimental Brain Research},
  volume    = {7},
  number    = {2},
  pages     = {158--168},
  year      = {1969},
  doi       = {10.1007/BF00235441},
  publisher = {Springer}
}

@article{plopski2022eye,
author = {Plopski, Alexander and Hirzle, Teresa and Norouzi, Nahal and Qian, Long and Bruder, Gerd and Langlotz, Tobias},
title = {The Eye in Extended Reality: A Survey on Gaze Interaction and Eye Tracking in Head-worn Extended Reality},
year = {2022},
issue_date = {March 2023},
publisher = {Association for Computing Machinery},
address = {New York, NY, USA},
volume = {55},
number = {3},
issn = {0360-0300},
url = {https://doi.org/10.1145/3491207},
doi = {10.1145/3491207},
journal = {ACM Comput. Surv.},
month = mar,
articleno = {53},
numpages = {39}
}

@article{niehorster2020slippage,
  title     = {The impact of slippage on the data quality of head-worn eye trackers},
  author    = {Niehorster, Diederick C. and Santini, Thiago and Hessels, Roy S. and Hooge, Ignace T. C. and Kasneci, Enkelejda and Nystr{\"o}m, Marcus},
  journal   = {Behavior Research Methods},
  volume    = {52},
  pages     = {1140--1160},
  year      = {2020},
  month     = jan,
  doi       = {10.3758/s13428-019-01307-0},
  publisher = {Springer}
}

@inproceedings{reiter2022look,
author = {Reiter, Katharina and Pfeuffer, Ken and Esteves, Augusto and Mittermeier, Tim and Alt, Florian},
title = {Look \& Turn: One-handed and Expressive Menu Interaction by Gaze and Arm Turns in VR},
year = {2022},
isbn = {9781450392525},
publisher = {Association for Computing Machinery},
address = {New York, NY, USA},
url = {https://doi.org/10.1145/3517031.3529233},
doi = {10.1145/3517031.3529233},
booktitle = {2022 Symposium on Eye Tracking Research and Applications},
articleno = {66},
numpages = {7},
location = {Seattle, WA, USA},
series = {ETRA '22}
}

@article{nwagu2023eeg,
author = {Nwagu, Chukwuemeka and AlSlaity, Alaa and Orji, Rita},
title = {EEG-Based Brain-Computer Interactions in Immersive Virtual and Augmented Reality: A Systematic Review},
year = {2023},
issue_date = {June 2023},
publisher = {Association for Computing Machinery},
address = {New York, NY, USA},
volume = {7},
number = {EICS},
url = {https://doi.org/10.1145/3593226},
doi = {10.1145/3593226},
journal = {Proc. ACM Hum.-Comput. Interact.},
month = jun,
articleno = {174},
numpages = {33}
}

@article{kotani2025stimulus,
  title={Stimulus-preceding negativity and P3 reflecting relative amounts of attention allocation between primary and secondary tasks},
  author={Kotani, Yasunori and Horai, Atsushi and Ohgami, Yoshimi and Gheorghe, Lucian},
  journal={NeuroReport},
  pages={10--1097},
  year={2025},
  publisher={LWW}
}

@inproceedings{penkar2012designing,
author = {Penkar, Abdul Moiz and Lutteroth, Christof and Weber, Gerald},
title = {Designing for the eye: design parameters for dwell in gaze interaction},
year = {2012},
isbn = {9781450314381},
publisher = {Association for Computing Machinery},
address = {New York, NY, USA},
url = {https://doi.org/10.1145/2414536.2414609},
doi = {10.1145/2414536.2414609},
booktitle = {Proceedings of the 24th Australian Computer-Human Interaction Conference},
pages = {479–488},
numpages = {10},
location = {Melbourne, Australia},
series = {OzCHI '12}
}

@article{narkar2024gazeintent,
author = {Narkar, Anish S. and Michalak, Jan J. and Peacock, Candace E. and David-John, Brendan},
title = {GazeIntent: Adapting Dwell-time Selection in VR Interaction with Real-time Intent Modeling},
year = {2024},
issue_date = {May 2024},
publisher = {Association for Computing Machinery},
address = {New York, NY, USA},
volume = {8},
number = {ETRA},
url = {https://doi.org/10.1145/3655600},
doi = {10.1145/3655600},
journal = {Proc. ACM Hum.-Comput. Interact.},
month = may,
articleno = {226},
numpages = {18}
}

@inproceedings{starker1990gaze,
author = {Starker, India and Bolt, Richard A.},
title = {A gaze-responsive self-disclosing display},
year = {1990},
isbn = {0201509326},
publisher = {Association for Computing Machinery},
address = {New York, NY, USA},
url = {https://doi.org/10.1145/97243.97245},
doi = {10.1145/97243.97245},
booktitle = {Proceedings of the SIGCHI Conference on Human Factors in Computing Systems},
pages = {3–10},
numpages = {8},
location = {Seattle, Washington, USA},
series = {CHI '90}
}

@article{yan2020head,
author = {Yan, Yukang and Shi, Yingtian and Yu, Chun and Shi, Yuanchun},
title = {HeadCross: Exploring Head-Based Crossing Selection on Head-Mounted Displays},
year = {2020},
issue_date = {March 2020},
publisher = {Association for Computing Machinery},
address = {New York, NY, USA},
volume = {4},
number = {1},
url = {https://doi.org/10.1145/3380983},
doi = {10.1145/3380983},
journal = {Proc. ACM Interact. Mob. Wearable Ubiquitous Technol.},
month = mar,
articleno = {35},
numpages = {22}
}

@inproceedings{gabel2024guiding,
author = {Gabel, Jenny and Schmidt, Susanne and Pfeuffer, Ken and Steinicke, Frank},
title = {Guiding Handrays in Virtual Reality: Comparison of Gaze- and Object-Based Assistive Raycast Redirection},
year = {2024},
isbn = {9798400710889},
publisher = {Association for Computing Machinery},
address = {New York, NY, USA},
url = {https://doi.org/10.1145/3677386.3682080},
doi = {10.1145/3677386.3682080},
booktitle = {Proceedings of the 2024 ACM Symposium on Spatial User Interaction},
articleno = {27},
numpages = {12},
location = {Trier, Germany},
series = {SUI '24}
}

@inproceedings{gabel2023redirecting,
author = {Gabel, Jenny and Schmidt, Susanne and Ariza, Oscar and Steinicke, Frank},
title = {Redirecting Rays: Evaluation of Assistive Raycasting Techniques in Virtual Reality},
year = {2023},
isbn = {9798400703287},
publisher = {Association for Computing Machinery},
address = {New York, NY, USA},
url = {https://doi.org/10.1145/3611659.3615716},
doi = {10.1145/3611659.3615716},
booktitle = {Proceedings of the 29th ACM Symposium on Virtual Reality Software and Technology},
articleno = {38},
numpages = {11},
location = {Christchurch, New Zealand},
series = {VRST '23}
}

@inproceedings{chen2023gaze,
author = {Chen, Di Laura and Giordano, Marcello and Benko, Hrvoje and Grossman, Tovi and Santosa, Stephanie},
title = {GazeRayCursor: Facilitating Virtual Reality Target Selection by Blending Gaze and Controller Raycasting},
year = {2023},
isbn = {9798400703287},
publisher = {Association for Computing Machinery},
address = {New York, NY, USA},
url = {https://doi.org/10.1145/3611659.3615693},
doi = {10.1145/3611659.3615693},
booktitle = {Proceedings of the 29th ACM Symposium on Virtual Reality Software and Technology},
articleno = {19},
numpages = {11},
location = {Christchurch, New Zealand},
series = {VRST '23}
}

@ARTICLE{tian2025amplitude,
  author={Tian, Yang and Zhang, Youpeng and Yan, Yukang and Zhao, Shengdong and Ma, Xiaojuan and Shi, Yuanchun},
  journal={IEEE Transactions on Visualization and Computer Graphics}, 
  title={AmplitudeArrow: On-the-Go AR Menu Selection Using Consecutive Simple Head Gestures and Amplitude Visualization}, 
  year={2025},
  volume={},
  number={},
  pages={1-15},
  doi={10.1109/TVCG.2025.3531378}
}

@inproceedings{hendrikson2020head,
author = {Henrikson, Rorik and Grossman, Tovi and Trowbridge, Sean and Wigdor, Daniel and Benko, Hrvoje},
title = {Head-Coupled Kinematic Template Matching: A Prediction Model for Ray Pointing in VR},
year = {2020},
isbn = {9781450367080},
publisher = {Association for Computing Machinery},
address = {New York, NY, USA},
url = {https://doi.org/10.1145/3313831.3376489},
doi = {10.1145/3313831.3376489},
booktitle = {Proceedings of the 2020 CHI Conference on Human Factors in Computing Systems},
pages = {1–14},
numpages = {14},
location = {Honolulu, HI, USA},
series = {CHI '20}
}

@article{chwilla1991event,
  title={Event-related potential correlates of non-motor anticipation},
  author={Chwilla, Dorothee J and Brunia, Cornelis HM},
  journal={Biological Psychology},
  volume={32},
  number={2-3},
  pages={125--141},
  year={1991},
  publisher={Elsevier}
}

@article{koulieris2019neareye,
  title     = {Near-Eye Display and Tracking Technologies for Virtual and Augmented Reality},
  author    = {Koulieris, George A. and Ak{\c{s}}it, Kaan and Stengel, Michael and Mantiuk, Rafa{\l} K. and Mania, Katerina and Richardt, Christian},
  journal   = {Computer Graphics Forum},
  volume    = {38},
  number    = {2},
  pages     = {493--519},
  year      = {2019},
  month     = jun,
  doi       = {10.1111/cgf.13654},
  publisher = {Wiley}
}

@inproceedings{he2024pupil,
  title={Pupil size reflects the relevance of reward prediction error and estimation uncertainty in upcoming choice},
  author={He, Zoe W and L'H{\^o}tellier, Ma{\"e}va and Paunov, Alexander and Guo, Dalin and Meyniel, Florent and Yu, Angela J},
  booktitle={Proceedings of the Annual Meeting of the Cognitive Science Society},
  volume={46},
  year={2024}
}

@article{fan2023pupil,
  title={Pupil size encodes uncertainty during exploration},
  author={Fan, Haoxue and Burke, Taylor and Sambrano, Deshawn Chatman and Dial, Emily and Phelps, Elizabeth A and Gershman, Samuel J},
  journal={Journal of cognitive neuroscience},
  volume={35},
  number={9},
  pages={1508--1520},
  year={2023},
  publisher={MIT Press One Broadway, 12th Floor, Cambridge, Massachusetts 02142, USA~…}
}

@article{studer2016psychophysiological,
  title={Psychophysiological arousal and inter-and intraindividual differences in risk-sensitive decision making},
  author={Studer, Bettina and Scheibehenne, Benjamin and Clark, Luke},
  journal={Psychophysiology},
  volume={53},
  number={6},
  pages={940--950},
  year={2016},
  publisher={Wiley Online Library}
}

@article{wichary2016probabilistic,
  title={Probabilistic inferences under emotional stress: how arousal affects decision processes},
  author={Wichary, Szymon and Mata, Rui and Rieskamp, J{\"o}rg},
  journal={Journal of Behavioral Decision Making},
  volume={29},
  number={5},
  pages={525--538},
  year={2016},
  publisher={Wiley Online Library}
}

@article{eraslan2015eye,
  title={Eye tracking scanpath analysis techniques on web pages: A survey, evaluation and comparison},
  author={Eraslan, Sukru and Yesilada, Yeliz and Harper, Simon},
  journal={Journal of Eye Movement Research},
  volume={9},
  number={1},
  pages={2},
  year={2015},
  publisher={Bern Open Publishing},
  doi={10.16910/jemr.9.1.2}

}

@article{wollstadt2021quantifying,
  title={Quantifying the predictability of visual scanpaths using active information storage},
  author={Wollstadt, Patricia and Hasenj{\"a}ger, Martina and Wiebel-Herboth, Christiane B},
  journal={Entropy},
  volume={23},
  number={2},
  pages={167},
  year={2021},
  publisher={MDPI},
  doi={10.3390/e23020167}
}

@article{posner1980orienting,
  title={Orienting of attention},
  author={Posner, Michael I},
  journal={Quarterly Journal of Experimental Psychology},
  volume={32},
  number={1},
  pages={3--25},
  year={1980},
  publisher={Taylor \& Francis},
  doi={10.1080/00335558008248231}
}

@article{iacoboni1999cortical,
  title={Cortical mechanisms of human imitation},
  author={Iacoboni, Marco and Woods, Roger P and Brass, Marcel and Bekkering, Harold and Mazziotta, John C and Rizzolatti, Giacomo},
  journal={Science},
  volume={286},
  number={5449},
  pages={2526--2528},
  year={1999},
  publisher={American Association for the Advancement of Science},
  doi={10.1126/science.286.5449.2526}
}

@article{sakai2008task,
  title={Task set and prefrontal cortex},
  author={Sakai, Katsuyuki},
  journal={Annual Review of Neuroscience},
  volume={31},
  pages={219--245},
  year={2008},
  publisher={Annual Reviews},
  doi={10.1146/annurev.neuro.31.060407.125642}
}

@article{kastner2000mechanisms,
  title={Mechanisms of visual attention in the human cortex},
  author={Kastner, Sabine and Ungerleider, Leslie G},
  journal={Annual Review of Neuroscience},
  volume={23},
  number={1},
  pages={315--341},
  year={2000},
  publisher={Annual Reviews},
  doi={10.1146/annurev.neuro.23.1.315}
}

@article{vanboxtel2004cortical,
  title={Cortical measures of anticipation},
  author={van Boxtel, Geert JM and B{\"o}cker, Koen BE},
  journal={Journal of Psychophysiology},
  volume={18},
  number={2-3},
  pages={61--76},
  year={2004},
  publisher={Hogrefe \& Huber Publishers},
  doi={10.1027/0269-8803.18.23.61}
}

@article{friston2009predictive,
  title={Predictive coding under the free-energy principle},
  author={Friston, Karl and Kiebel, Stefan},
  journal={Philosophical Transactions of the Royal Society B: Biological Sciences},
  volume={364},
  number={1521},
  pages={1211--1221},
  year={2009},
  publisher={The Royal Society},
  doi={10.1098/rstb.2008.0300}
}

@article{catena2012brain,
  title={The brain network of expectancy and uncertainty processing},
  author={Catena, Andr{\'e}s and Perales, Jos{\'e} C and Meg{\'\i}as, Alberto and C{\'a}ndido, Antonio and Jara, Eva and Maldonado, Antonio},
  journal={PloS One},
  volume={7},
  number={7},
  pages={e40252},
  year={2012},
  publisher={Public Library of Science San Francisco, USA},
  doi={10.1371/journal.pone.0040252}
}

@article{walentowska2018relevance,
  title={Relevance and uncertainty jointly influence reward anticipation at the level of the {SPN} {ERP} component},
  author={Walentowska, Wioleta and Paul, Katharina and Severo, Mario Carlo and Moors, Agnes and Pourtois, Gilles},
  journal={International Journal of Psychophysiology},
  volume={132},
  pages={287--297},
  year={2018},
  publisher={Elsevier},
  doi={10.1016/j.ijpsycho.2017.11.005}
}

@article{clark2013whatever,
  title={Whatever next? {P}redictive brains, situated agents, and the future of cognitive science},
  author={Clark, Andy},
  journal={Behavioral and Brain Sciences},
  volume={36},
  number={3},
  pages={181--204},
  year={2013},
  publisher={Cambridge University Press},
  doi={10.1017/S0140525X12000477}
}

@inproceedings{ahuja2021pose,
  title={Pose-on-the-go: Approximating user pose with smartphone sensor fusion and inverse kinematics},
  author={Ahuja, Karan and Mayer, Sven and Goel, Mayank and Harrison, Chris},
  booktitle={Proceedings of the 2021 CHI Conference on Human Factors in Computing Systems},
  pages={1--12},
  year={2021}
}

@article{feldman2010attention,
  title={Attention, uncertainty, and free-energy},
  author={Feldman, Harriet and Friston, Karl J},
  journal={Frontiers in human neuroscience},
  volume={4},
  pages={215},
  year={2010},
  publisher={Frontiers Research Foundation}
}

@article{alexander2016readiness,
  title={Readiness potentials driven by non-motoric processes},
  author={Alexander, Patrick and Schlegel, Alexander and Sinnott-Armstrong, Walter and Roskies, Adina L and Wheatley, Thalia and Tse, Peter U},
  journal={Consciousness and Cognition},
  volume={39},
  pages={38--47},
  year={2016},
  publisher={Elsevier},
  doi={10.1016/j.concog.2015.11.011}
}

@article{schurger2021readiness,
  title={What is the readiness potential?},
  author={Schurger, Aaron and Hu, Pengbo and Pak, Joanna and Roskies, Adina L},
  journal={Trends in Cognitive Sciences},
  volume={25},
  number={7},
  pages={558--570},
  year={2021},
  publisher={Elsevier},
  doi={10.1016/j.tics.2021.04.001}
}

@article{megias2018electroencephalographic,
  title={Electroencephalographic evidence of abnormal anticipatory uncertainty processing in gambling disorder patients},
  author={Meg{\'\i}as, Alberto and Navas, Juan F and Perandr{\'e}s-G{\'o}mez, Antonia and Maldonado, Antonio and Catena, Andr{\'e}s and Perales, Jos{\'e} C},
  journal={Journal of Gambling Studies},
  volume={34},
  number={2},
  pages={321--338},
  year={2018},
  publisher={Springer},
  doi={10.1007/s10899-017-9693-3}
}

@article{shibasaki2006components,
  title={What is the Bereitschaftspotential?},
  author={Shibasaki, Hiroshi and Hallett, Mark},
  journal={Clinical Neurophysiology},
  volume={117},
  number={11},
  pages={2341--2356},
  year={2006},
  publisher={Elsevier}
}

@article{eimer1998lateralized,
  title={The lateralized readiness potential as an on-line measure of central response activation processes},
  author={Eimer, Martin},
  journal={Behavior Research Methods, Instruments, \& Computers},
  volume={30},
  number={1},
  pages={146--156},
  year={1998},
  publisher={Springer}
}

@article{seidel2014uncertainty,
  title={Uncertainty during pain anticipation: The adaptive value of preparatory processes},
  author={Seidel, Eva-Maria and Pfabigan, Daniela M and Hahn, Andreas and Sladky, Ronald and Grahl, Arvina and Paul, Katharina and Kraus, Christoph and K{\"u}blb{\"o}ck, Manfred and Kranz, Georg S and Hummer, Allan and others},
  journal={Human Brain Mapping},
  volume={36},
  number={2},
  pages={744--755},
  year={2015},
  publisher={Wiley Online Library}
}

@article{ang2012clinical,
  title={A large clinical study on the ability of stroke patients to use an EEG-based motor imagery brain-computer interface},
  author={Ang, Kai Keng and Guan, Cuntai and Chua, Karen Sui Geok and Ang, Beng Ti and Kuah, Christopher WK and Wang, Chuanchu and Phua, Kok Soon and Chin, Zhou Yong and Zhang, Haihong},
  journal={Clinical EEG and Neuroscience},
  volume={42},
  number={4},
  pages={253--258},
  year={2011},
  publisher={SAGE Publications Sage CA: Los Angeles, CA}
}

@incollection{majaranta2012communication,
  title={Communication and text entry by gaze},
  author={Majaranta, P{\"a}ivi},
  booktitle={Gaze interaction and applications of eye tracking: Advances in assistive technologies},
  pages={63--77},
  year={2012},
  publisher={IGI Global Scientific Publishing}
}

@inproceedings{ribeiro2016lime,
  title={“Why Should I Trust You?” Explaining the Predictions of Any Classifier},
  author={Ribeiro, Marco Tulio and Singh, Sameer and Guestrin, Carlos},
  booktitle={Proceedings of the 22nd ACM SIGKDD International Conference on Knowledge Discovery and Data Mining},
  pages={1135--1144},
  year={2016},
  publisher={ACM}
}

@article{rejer2025averagedlime,
  title={AveragedLIME for general explanations in the EEG domain},
  author={Rejer, Izabela and Gago, Izabela},
  journal={NeuroImage},
  volume={323},
  pages={121588},
  year={2025},
  publisher={Elsevier}
}

@inproceedings{xiao2020vibrocomm,
  author = {Xiao, Robert and Cao, Teng and Guo, Ning and Zhuo, Jiannan and Zhang, Yang and Harrison, Chris},
  title = {VibroComm: Using Vibroacoustics to Facilitate Communication Between Mobile Devices},
  booktitle = {Proceedings of the 2020 CHI Conference on Human Factors in Computing Systems},
  series = {CHI '20},
  year = {2020},
  pages = {1--12},
  doi = {10.1145/3313831.3376139},
  publisher = {Association for Computing Machinery},
  address = {New York, NY, USA}
}

@inproceedings{davidjohn2021towards,
  author = {David-John, Brendan and Peacock, Candace and Zhang, Ting and Murdison, T. Scott and Benko, Hrvoje and Jonker, Tanya R.},
  title = {Towards Gaze-Based Prediction of the Intent to Interact in Virtual Reality},
  booktitle = {ACM Symposium on Eye Tracking Research and Applications},
  series = {ETRA '21 Short Papers},
  year = {2021},
  articleno = {2},
  pages = {1--7},
  doi = {10.1145/3448018.3458008},
  publisher = {Association for Computing Machinery},
  address = {New York, NY, USA}
}

@inproceedings{mohan2018dualgaze,
  author = {Mohan, Pallavi and Goh, Weng-Hong and Mok, Chan-Wai and Fu, Chi-Wing},
  title = {DualGaze: Addressing the Midas Touch Problem in Gaze Mediated VR Interaction},
  booktitle = {2018 IEEE International Symposium on Mixed and Augmented Reality Adjunct (ISMAR-Adjunct)},
  year = {2018},
  pages = {79--84},
  doi = {10.1109/ISMAR-Adjunct.2018.00036},
  publisher = {IEEE},
  address = {Munich, Germany}
}

@inproceedings{wolf2021gaze,
  author = {Wolf, Julian and Lohmeyer, Quentin and Holz, Christian and Meboldt, Mirko},
  title = {Gaze Comes in Handy: Predicting and Preventing Erroneous Hand Actions in AR-Supported Manual Tasks},
  booktitle = {2021 IEEE International Symposium on Mixed and Augmented Reality (ISMAR)},
  year = {2021},
  pages = {180--189},
  doi = {10.1109/ISMAR52148.2021.00031},
  publisher = {IEEE},
  address = {Bari, Italy}
}

@inproceedings{zhao2025spatial,
  author = {Zhao, Yu and Bodenheimer, Bobby},
  title = {How Spatial Ability Affects Response to Gaze-Adaptive Cueing in Mixed Reality Spatial Navigation},
  booktitle = {2025 IEEE International Symposium on Mixed and Augmented Reality (ISMAR)},
  year = {2025},
  note = {Paper ID 2194, accepted for publication},
  publisher = {IEEE}
}

@article{severitt2025gaze,
  author = {Severitt, Bj{\"o}rn R. and Castner, Nora and Wahl, Siegfried},
  title = {The Interplay of User Preference and Precision in Different Gaze-Based Interaction Methods in Virtual Environments},
  journal = {Frontiers in Virtual Reality},
  volume = {6},
  pages = {1576962},
  year = {2025},
  month = {March},
  doi = {10.3389/frvir.2025.1576962},
  publisher = {Frontiers Media SA}
}

%
%

\appendix
\section{Appendix}

\subsection{Model Selection}

To identify the most appropriate random-effects structure, a series of nested linear mixed-effects models were compared using maximum likelihood estimation. Models differed in the inclusion of random intercepts for \textit{Order}, \textit{channel}, and \textit{Scene}, in addition to the baseline random intercept for \textit{PID}. Model fit was evaluated using Akaike’s Information Criterion (AIC), Bayesian Information Criterion (BIC), and likelihood ratio tests.

\begin{table}[H]
\centering
\caption{Model comparison for negative peak amplitude (maximum likelihood estimation).}
\label{tab:modelcomp}
\begin{tabular}{lrrrr}
\toprule
Random effects structure & AIC & BIC & logLik \\
\midrule
PID only                        & 509023.5 & 509078.6 & -254505.7 \\
PID + Scene                     & 509007.0 & 509071.3 & -254496.5 \\
PID + Trial                     & 508713.9 & 508778.2 & -254350.0 \\
PID + Trial + Scene             & 508696.5 & 508770.0 & -254340.3 \\
PID + Channel                   & 508473.5 & 508537.8 & -254229.7 \\
PID + Channel + Scene           & 508456.8 & 508530.3 & -254220.4 \\
PID + Trial + Channel           & 508160.6 & 508234.1 & -254072.3 \\
PID + Trial + Channel + Scene   & \textbf{508143.0} & \textbf{508225.6} & \textbf{-254062.5} \\
\bottomrule
\end{tabular}
\label{app:table}
\end{table}

Likelihood ratio tests confirmed that the full model (PID + Trial + Channel + Scene) provided a significantly better fit than simpler alternatives (all $p < .001$). Accordingly, this model was selected for all subsequent analyses.

\subsection{Neural Networks Saliency Maps}
To better understand how each neural network made its predictions, we generated LIME-based saliency maps for all five architectures. These visualizations highlight which EEG channels and time periods contributed most to the model’s decision for each trial. By averaging saliency values across correctly classified samples and participants, we obtain a stable picture of the neural features each architecture relies on. The following figures show both the temporal importance patterns and the corresponding scalp topographies for Select and Observe conditions, as well as their difference.

\begin{figure*}[t!]
    \centering
    \includegraphics[width=0.95\linewidth]{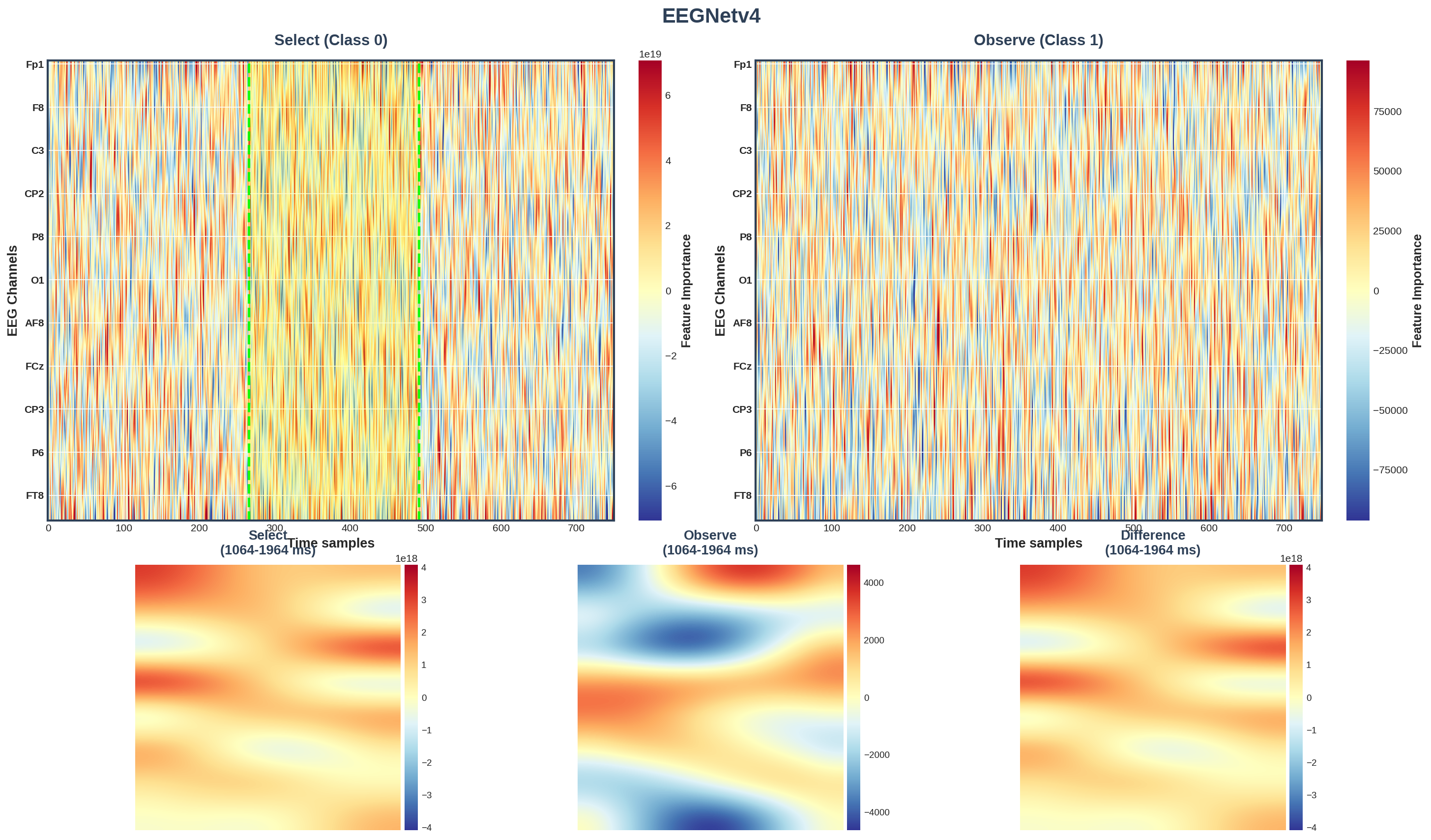}
    \caption{\textbf{LIME-based saliency maps for EEGNetv4.}
    Top panels show temporal heatmaps of LIME feature importance across all 64 EEG channels and 750 time samples for \emph{Select} (Class~0, left) and \emph{Observe} (Class~1, right). Green dashed lines and yellow shading indicate the automatically identified analysis window of maximal activation (1064--1964~ms). Bottom panels display topographic projections of averaged feature importance within this window for Select, Observe, and their Difference. Warm colors (red/orange) denote positive feature importance, cool colors (blue) negative importance. EEGNetv4 exhibits a focal centro-parietal cluster centered on \textbf{Pz, CPz, Cz}, extending into \textit{POz, PO3, PO4, O1, O2, Iz}. Spatial and temporal concentration are near ceiling ($H_{\mathrm{spatial}}=.99$, $H_{\mathrm{temporal}}=.98$) with negligible inter-trial variability ($SD<.001$), indicating stable and physiologically consistent reliance on posterior SPN-related preparatory components. Grand averages are computed across $N=28$ participants.}
    \label{fig:lime_eegnetv4}
\end{figure*}

\begin{figure*}[t!]
    \centering
    \includegraphics[width=0.95\linewidth]{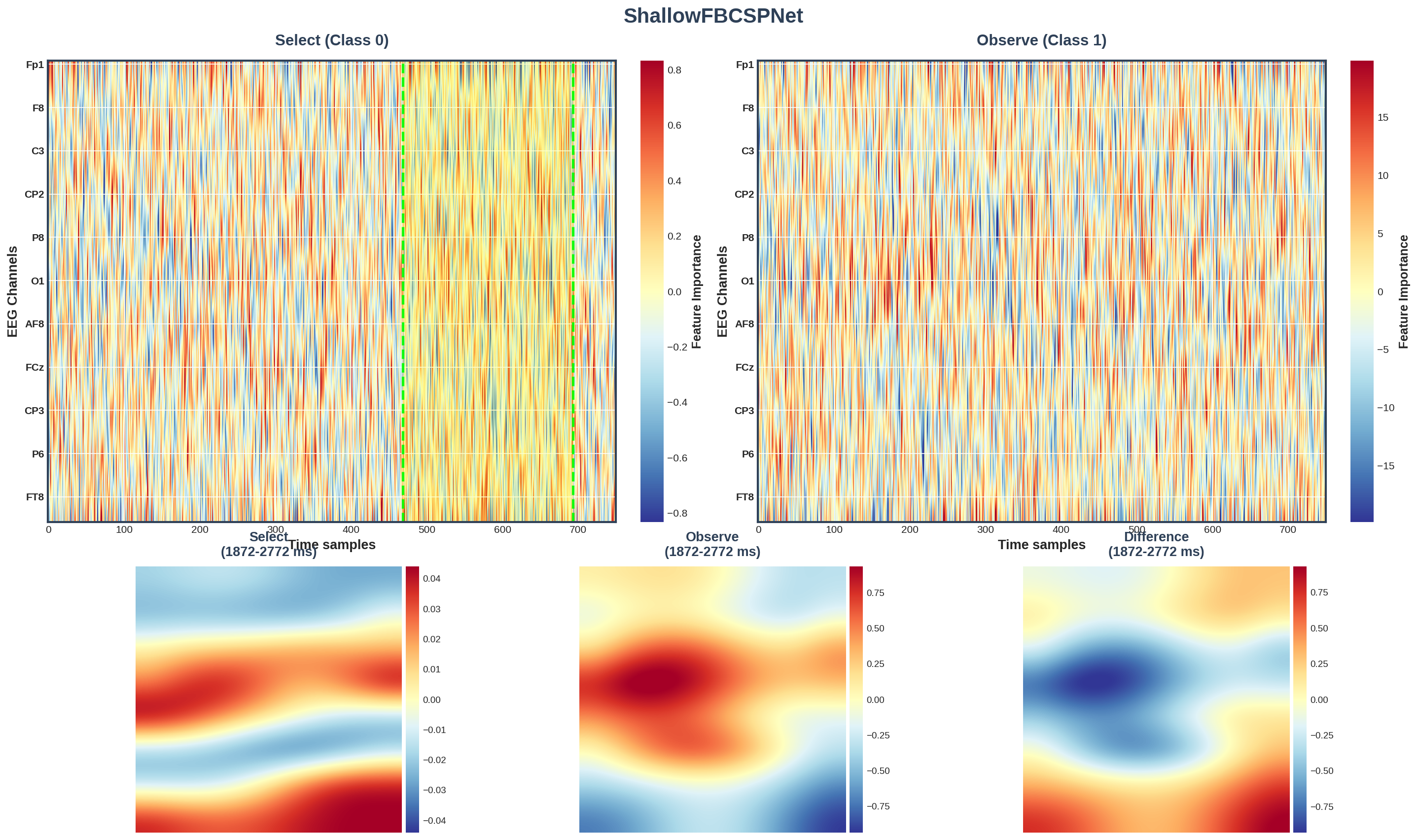}
    \caption{\textbf{LIME-based saliency maps for ShallowFBCSPNet.}
    Top panels depict temporal heatmaps of LIME feature importance for \emph{Select} (Class~0, left) and \emph{Observe} (Class~1, right) across 64 EEG channels and 750 time samples. The green dashed lines and yellow shading mark the automatically selected analysis window (1872--2772~ms) in which overall importance peaks. Bottom panels show the corresponding scalp topographies of averaged feature importance for Select, Observe, and their Difference. ShallowFBCSPNet produces a posterior pattern highly similar to EEGNetv4, with peak importance over \textit{CP1--CP4, P1--P4, P5/P6, PO3/PO4}. Concentration metrics remain extremely high ($H_{\mathrm{spatial}}=.99$, $H_{\mathrm{temporal}}=.99$) and inter-trial dispersion is minimal ($SD<.001$), suggesting that the model consistently emphasizes midline and adjacent parietal channels associated with slow negative SPN activity. Grand averages are computed across $N=28$ participants.}
    \label{fig:lime_shallow}
\end{figure*}

\begin{figure*}[t!]
    \centering
    \includegraphics[width=0.95\linewidth]{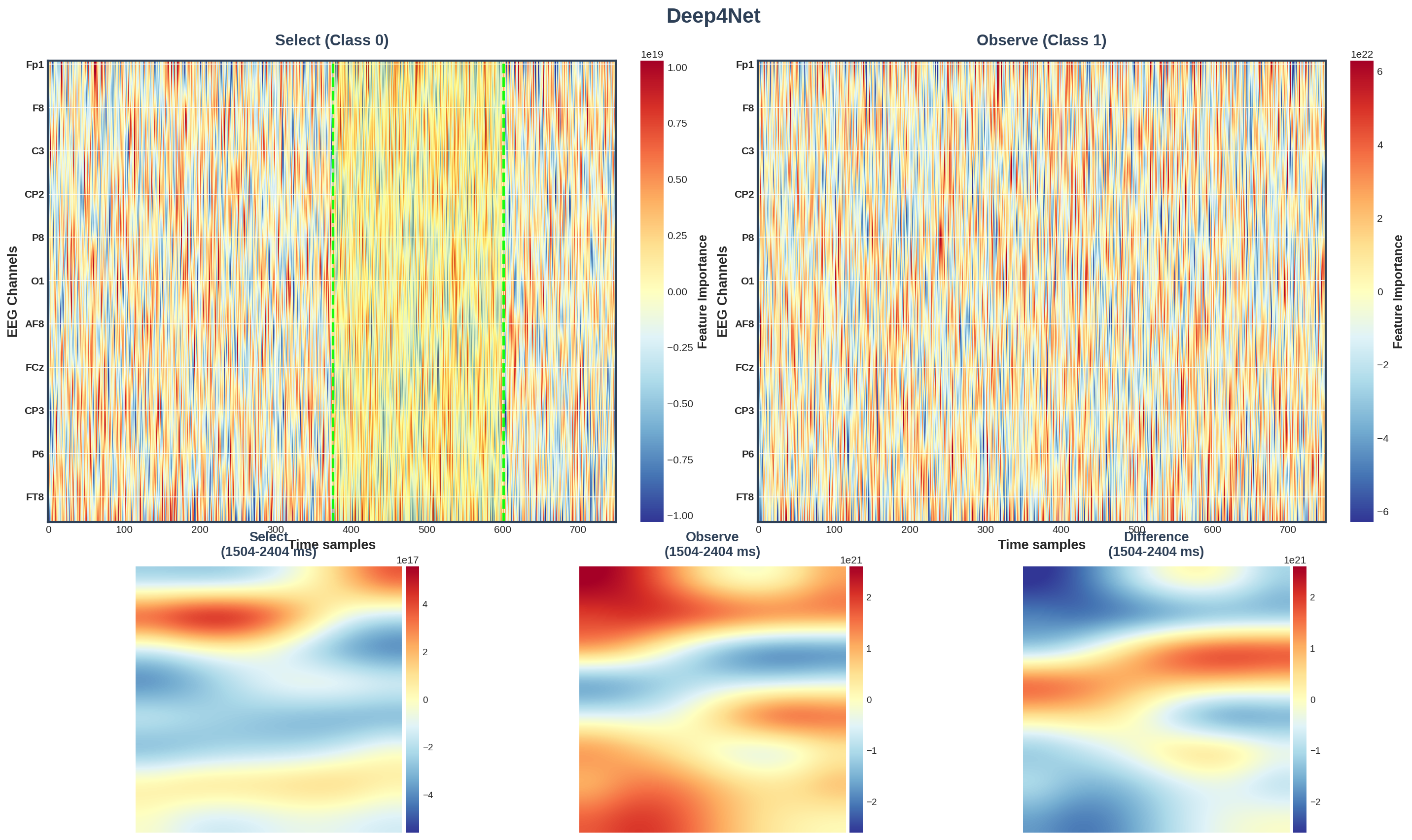}
    \caption{\textbf{LIME-based saliency maps for Deep4Net.}
    Top panels show temporal LIME feature-importance heatmaps for \emph{Select} (Class~0, left) and \emph{Observe} (Class~1, right) across 64 channels and 750 samples. The green dashed lines and yellow band highlight the automatically determined analysis window (1504--2404~ms) of maximal activation. Bottom panels present averaged topographic projections for Select, Observe, and their Difference over this window. Deep4Net exhibits the broadest yet still physiologically coherent posterior distribution, with strongest weights over \textit{Pz, CPz, P3/P4, P5/P6} and additional contributions from \textit{O1/O2} and \textit{POz}. Spatial and temporal concentration remain high ($H_{\mathrm{spatial}}=.98$, $H_{\mathrm{temporal}}=.98$) and inter-trial stability is excellent ($SD<.001$), indicating a consistent reliance on posterior SPN-like activity while allowing slightly more distributed weighting than the shallower architectures. Grand averages are computed across $N=28$ participants.}
    \label{fig:lime_deep4}
\end{figure*}

\begin{figure*}[t!]
    \centering
    \includegraphics[width=0.95\linewidth]{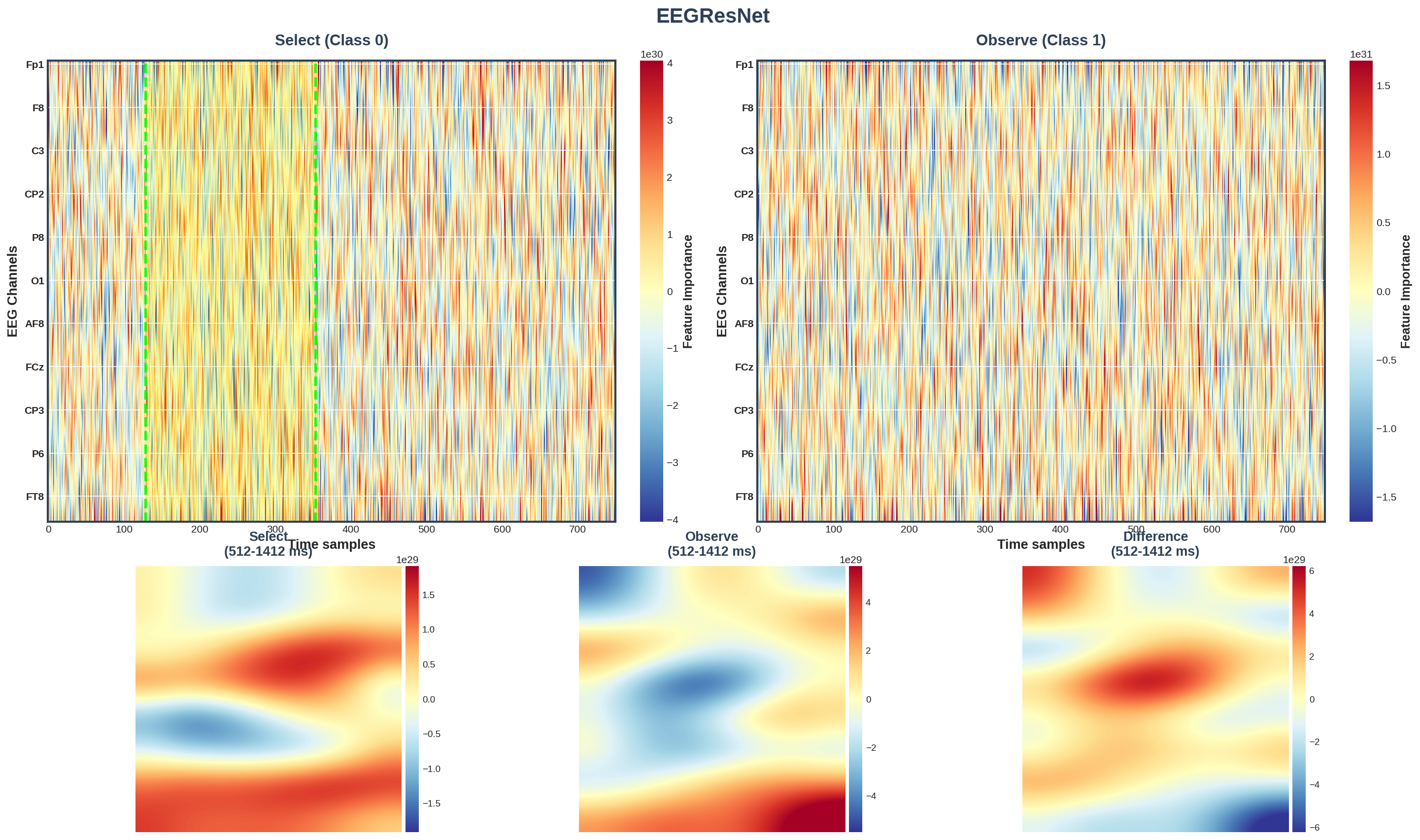}
    \caption{\textbf{LIME-based saliency maps for EEGResNet.}
    The upper panels display temporal heatmaps of LIME feature importance for \emph{Select} (Class~0, left) and \emph{Observe} (Class~1, right) across all channels and time samples. Green dashed lines and the yellow region indicate the automatically identified analysis period (512--1412~ms), capturing the earliest robust preparatory effects among the tested architectures. The lower panels show scalp maps of averaged feature importance for Select, Observe, and their Difference within this window. EEGResNet exhibits the most sharply localized pattern, with maximal saliency along the midline parietal axis \textit{CPz--Pz--POz} and secondary contributions at \textit{P1/P2} and \textit{PO3/PO4}. Both spatial and temporal concentration approach unity ($H_{\mathrm{spatial}}=.99$, $H_{\mathrm{temporal}}=.98$) with negligible variability across trials ($SD<.001$), indicating strongly convergent usage of SPN-related posterior features. Grand averages are computed across $N=28$ participants.}
    \label{fig:lime_resnet}
\end{figure*}

\begin{figure*}[t!]
    \centering
    \includegraphics[width=0.95\linewidth]{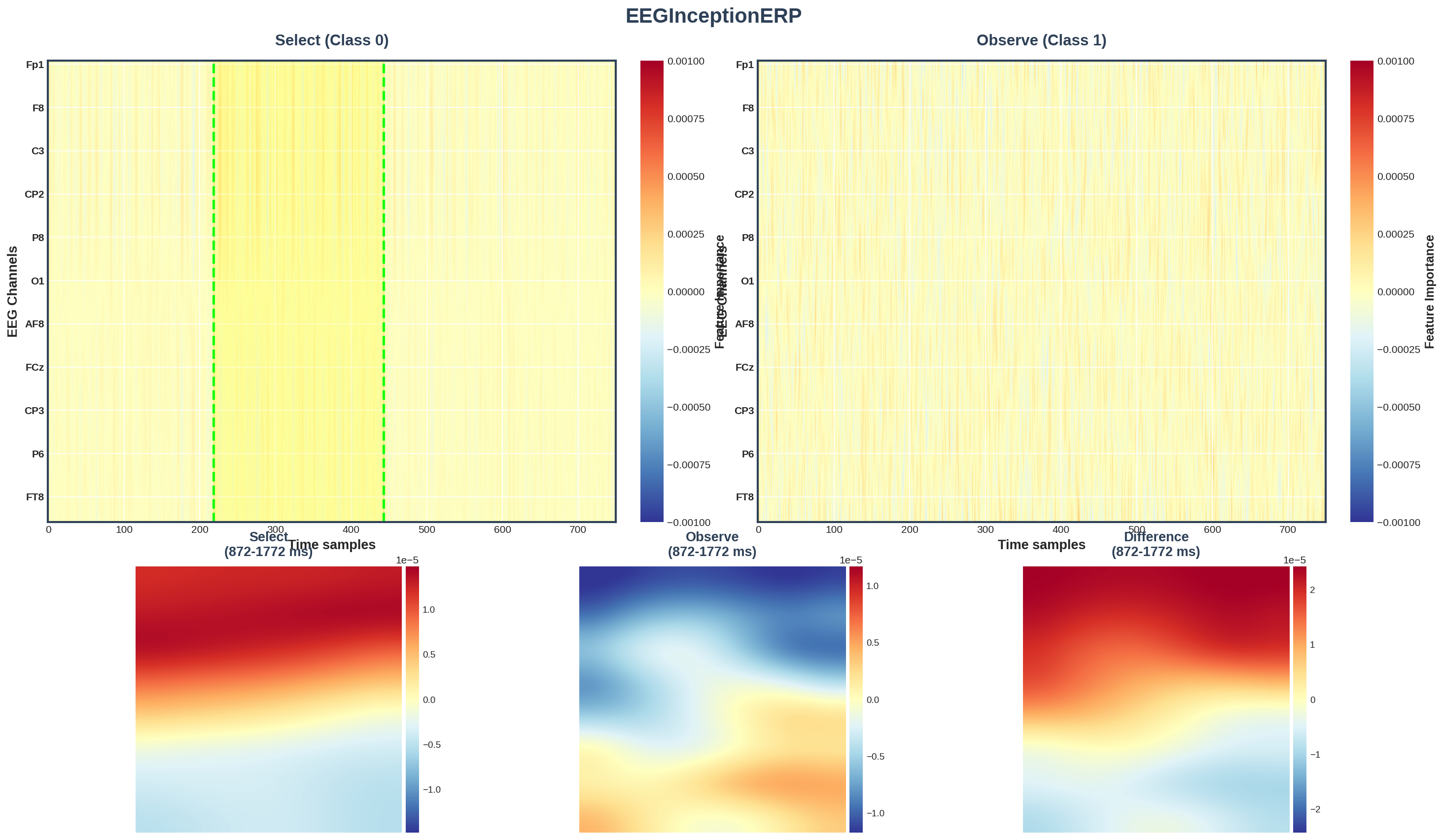}
    \caption{\textbf{LIME-based saliency maps for EEGInceptionERP.}
    Top panels illustrate temporal LIME feature-importance heatmaps for \emph{Select} (Class~0, left) and \emph{Observe} (Class~1, right) over all channels and time samples. The green dashed lines and yellow shading indicate the automatically selected analysis window (872--1772~ms), during which the model shows maximal sensitivity. Bottom panels provide topographic projections of averaged feature importance for Select, Observe, and their Difference within this window. EEGInceptionERP reveals a characteristic parietal--occipital ring with lateral extensions, with peak importance at \textit{PO7/PO8, PO3/PO4, P5/P6, O1/O2, Iz}. As with the other architectures, entropy-based concentration is extremely high ($H>.99$) and explanation stability is strong ($SD<.001$). The model’s weighting of late SPN time points aligns with its high temporal filter resolution, highlighting sustained preparatory activity over posterior scalp sites. Grand averages are computed across $N=28$ participants.}
    \label{fig:lime_inception}
\end{figure*}

%
%
\end{document}